\documentclass[twocolumn,pra,showpacs]{revtex4}
\usepackage{amsmath}
\usepackage{color}
\usepackage[pdftex]{graphicx}
\usepackage[caption=false]{subfig}
\usepackage{hyperref}
\newcommand{\ud}{\,\mathrm{d}}
\newcommand{\K}{\mathbf{K_{0}}}
\newcommand{\F}{F\left(\Delta/t\right)}
\begin{document}
  \title{Formation and decay of Bose-Einstein condensates in an excited band of a  double-well optical lattice}
  \author{Saurabh Paul}
  \affiliation{Joint Quantum Institute and University of Maryland, Maryland 20742, USA}
  \author{Eite Tiesinga}
  \affiliation{Joint Quantum Institute, National Institute of Standards and Technology and University of Maryland, Gaithersburg, Maryland 20899, USA}
  \date{\today}

  \begin{abstract}
    We study the formation and collision-aided decay of an ultra-cold atomic Bose-Einstein condensate in the first excited band of a double-well 2D-optical lattice with weak harmonic confinement in the perpendicular $z$ direction. This lattice geometry is based on an experiment by Wirth {\it et al.}\cite{wirth_evidence_2010}.
The double well is asymmetric, with the local ground state in the shallow well nearly degenerate with the first excited state of the adjacent deep well. We compare the band structure obtained from a tight-binding model with that obtained numerically using a plane wave basis. We find the tight binding model to be in quantitative agreement for the lowest two bands, qualitative for next two bands, and inadequate for even higher excited bands. The band widths of the excited bands are much larger than the harmonic oscillator energy spacing in the $z$ direction. We then study the thermodynamics of a non-interacting Bose gas in the first excited band. We estimate the condensate fraction and critical temperature, $T_c$, as functions of lattice parameters. For typical atom numbers, the critical energy $k_BT_c$, with $k_B$ the Boltzmann constant, is larger than the excited band widths and harmonic oscillator energy. Using conservation of total energy and atom number, we show that the temperature increases after the lattice transformation. Finally, we estimate the time scale for a two-body collision-aided  decay of the condensate as a function of lattice parameters. The decay involves two processes, the dominant one in which both colliding atoms decay to the ground band, and the second involving excitation of one atom to a higher band. For this estimate, we have used tight binding wave functions for the lowest four bands, and numerical estimates for higher bands. The decay rate rapidly increases with lattice depth, but close to the critical temperature, it stays  smaller than the tunneling rate between the $s$ and $p$ orbitals in adjacent wells.
  \end{abstract} 
  \maketitle

  \begin{section}{Introduction}\label{Introduction}
 Bose-Einstein condensation of ultracold atoms in the lowest Bloch band of a periodic optical-lattice potential \cite{bloch_many-body_2008, castin_bose-einstein_2001} occurs at zero quasi-momentum. This condensation can be understood from a bosonic Hubbard model. By varying the relative strength of tunneling energy and atom-atom  interaction strength, this model predicts a quantum phase transition between a delocalized compressible superfluid state and a localized incompressible Mott state \cite{jaksch_cold_1998, van_oosten_quantum_2001}. This was first observed by Greiner {\it et al.}\cite{greiner_quantum_2002}.

According to Feynman's ``no-node'' theorem \cite{Feynman}, the  zero quasi momentum ground state wave function is positive definite and has time-reversal (TR) symmetry. However, the no-node theorem does not apply to excited bands and the lowest energy state within the band may have non-zero quasi momenta. Thus, we can obtain exotic states of bosons with complex valued wave functions that spontaneously break the TR symmetry\cite{stojanovic_incommensurate_2008, wu_unconventional_2009}. Simultaneously, depending on the lattice parameters, two or more bands can become nearly degenerate and multiflavor Hubbard models can be realized where the atom-atom interaction strength can become of the order of bandgaps. In addition, for these bands with nearly degenerate excited orbitals, tunneling can be ``anisotropic'' in that different orbitals tunnel preferably along different primitive lattice vectors. This has lead to several proposals of unconventional Bose-Einstein condensates in optical lattices and predictions of rich orbital physics in higher bands \cite{isacsson_multiflavor_2005, liu_atomic_2006, stojanovic_incommensurate_2008, wu_pxy-orbital_2008, cai_complex_2011, wu_unconventional_2009}. For example, Wirth {\it et al.} \cite{wirth_evidence_2010} explores orbital superfluidity in $sp-$hybridized orbital bands in which condensation occurs at non-zero quasi momentum at the edge of the first Brillouin zone. Interestingly Refs.~\cite{isacsson_multiflavor_2005, stojanovic_incommensurate_2008} showed that Bose-Einstein condensates formed in excited Bloch bands may have very long lifetimes. Recently, such condensates have  been experimentally observed \cite{muller_state_2007, sebby-strabley_lattice_2006, wirth_evidence_2010, olschlager_unconventional_2011}.  

Optical lattices of two or more dimensions can have band crossings either at zero or non-zero quasi momentum. The band crossing at zero quasi momentum must occur between excited bands by virtue of the no-node theorem while the crossing at non-zero quasi momentum can occur between the ground and excited band. An example of the latter case occurs for a two-dimensional hexagonal lattice structure such as seen in Dirac cones of graphene \cite{tarruell_creating_2012, novoselov_two-dimensional_2005}. In fact, more than two bands can intersect \cite{wirth_evidence_2010}. The topological significance of band crossing points is now an area of active research \cite{sun_topological_2011,  olschlager_topologically_2012} (and references therein).

Optical lattices can be changed in real time and are thus tunable. The simplest example of this was the observation of the superfluid to Mott insulator transition of a cubic lattice by adiabatically increasing the lattice depth \cite{greiner_quantum_2002}. More advanced examples are the ability to dynamically change lattice geometries from triangular to hexagonal \cite{jo_ultracold_2012, tarruell_creating_2012}. Independent tunability of the sign of the nearest-neighbor tunneling energies has also been demonstrated \cite{PhysRevLett.99.220403, struck_quantum_2011}. Finally, we note that this tunability can enable us to study Berry's phases \cite{berry_quantal_1984} when band crossing of ground and/or excited bands occur for a particular choice of lattice depth and geometry. Adiabatic change of lattice parameters around this degeneracy point induce such phases.

   In this paper, we discuss the formation of a Bose-Einstein condensate in a 2D optical lattice, with a weak confinement along the $z$ axis.  Our paper is directly motivated by the experiment of Wirth \textit{et al.} \cite{wirth_evidence_2010}, in which Bose condensation is observed in the $p$ band of a quasi 2D double-well optical lattice. The lattice in the $xy$ plane consists of a checkerboard pattern of alternate deep and shallow wells, whose relative well depth can be controlled in real time by changing the phase difference $\theta$ between the counter-propagating lasers forming the optical lattice. Exciting atoms to higher bands is a two step process. Initially, as shown in Fig.~\ref{step1-a}, an angle $\theta$ is chosen such that one of the wells is much deeper than the other, and the atoms are largely confined to the local ground state of these deep wells. There is very little tunneling between the wells and the atoms form an array of one dimensional tubes along the $z$ axis.

The angle $\theta$ is now rapidly changed such that the ground $s$ state becomes degenerate or nearly degenerate with the first excited $p$ state of adjacent wells, which are now the deeper wells as shown in Fig.~\ref{step2-b}.  This change must be fast with respect to tunneling energies between adjacent wells, but adiabatic or slow with respect to the onsite energies of the well to which the atoms have been confined. After the change, the atoms can tunnel and populate the excited bands of the optical lattice. We thus observe a transition from a quasi one dimensional geometry to a three dimensional one, with atoms initially confined to  one dimensional wells getting distributed over the entire 2D lattice due to tunneling and elastic collision. Wirth \textit{et al} \cite{wirth_evidence_2010} saw that in this process, a Bose condensation is formed in the quasi momenta of lowest energy in the (first) excited band of the optical lattice. Interestingly, these quasi momenta are at the edge of the first Brillouin zone.

To model this physics, we perform a numerically exact as well as a tight binding calculation of the single particle band structure. A comparison shows that for the first four bands, the tight binding model gives a sufficient description of the band structure. We then performed a  calculation for the thermodynamic quantities for both cases in Fig.~\ref{fig:step12}, assuming that the atoms do not interact. In particular, we estimate the critical atom number  and critical temperature at which a Bose-Einstein condensate appears. We find that these quantities crucially depend on the detuning between the closely resonant $s$ and $p$ orbitals in adjacent wells.  We also justify the assumption of non-interacting bosons. 

Simultaneously, these atoms undergo collisional de-excitation to the ground band. De-excitation occurs when one or more atoms make a transition to the ground band in the $xy$ plane. This energy release is accompanied by an increase in energy along the $z$ direction. An example of such a process is shown in Fig.~\ref{step2-b}.
We will show that the dominant decay process is that one in which both colliding atoms decay to the ground band. We estimate the rate for this transition in the tight binding model as well as numerically, and find qualitative agreement. We provide analytical expression for the rate as a function of lattice depth and $\theta$.
With increasing lattice depth, processes where one of the colliding atoms decays to the ground band, and the other to an  excited band become important. These rates can only be evaluated numerically, as the tight binding model is inadequate for multiply-excited bands. At lattice parameters where adjacent $s$ and $p$ orbitals are resonant, the total decay rate is significantly smaller than the tunneling time scales, consistent with the  observations of the experiment \cite{wirth_evidence_2010}.

The paper is set up as follows. In Section~\ref{Hamiltonian}, we set up the Hamiltonian.  In Section~\ref{TB-model}  we discuss the tight binding calculation of the band structure followed by a numerical estimate of the same in Section~\ref{numerical-BS}. We also compare both the models in this section. Section \ref{Thermodynamics} uses the tight binding results to estimate the thermodynamic properties of the condensate. In section~\ref{lattice transformation} we study the lattice transformation used to excite atoms to the excited bands, and estimate the relationship between the initial and final temperatures. Section~\ref{lifetime} deals with the interaction processes that lead to the decay of the condensate. We conclude in Section\ref{Conclusion}.

     \begin{figure}
       \centering
       \subfloat[Part 1][]{\includegraphics[width=1.65in]{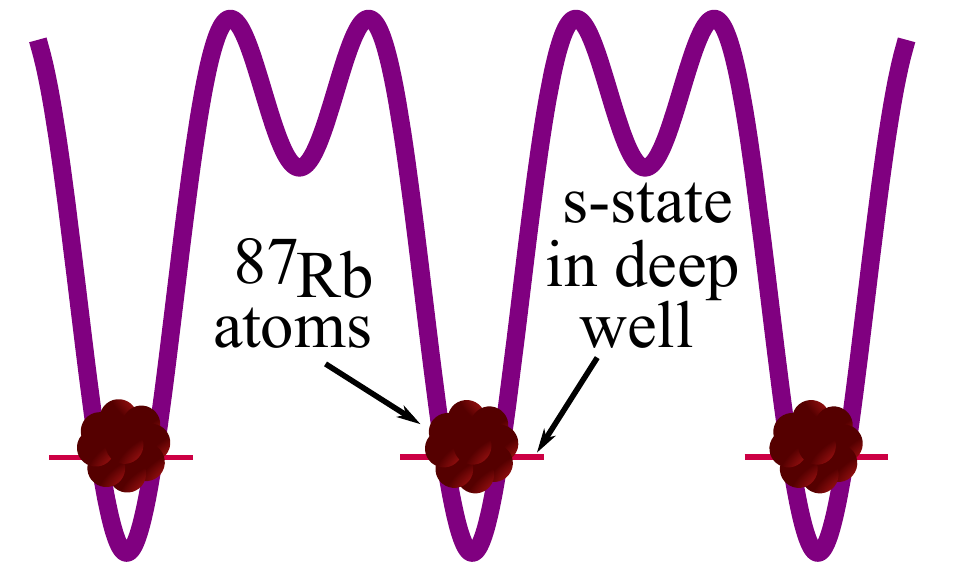} \label{step1-a}}
       \subfloat[part 2][]{\includegraphics[width=1.65in]{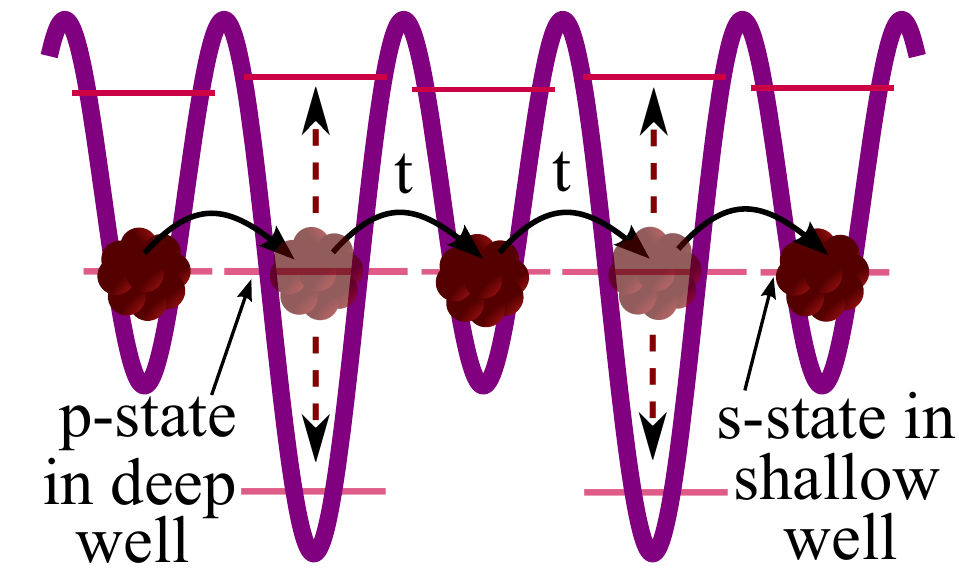} \label{step2-b}}
       \caption{(Color online) A one-dimensional schematic of lattice transformation to populate atoms in excited bands. Panel (a) shows the initial lattice configuration with atoms confined to the ground $s$ state of the deeper wells. The atoms form an array of one dimensional tubes where tunneling between adjacent tubes is negligible. Panel (b) shows the final lattice configuration, with $\theta$ chosen such that the $s$ state in the shallow well is nearly degenerate to the $p$ state of the deeper well. The well that was initially the deepest is now the shallower of the two. The hopping energy $t$ now distributes the atoms in the  excited bands of the 2D optical lattice. The vertical dashed lines represent a collisional decay process in which one of the atom decays to the ground band, while the other jumps to an excited band.
The decay processes largely involve vertical atom transitions within the deep sites.}
       \label{fig:step12}
     \end{figure}
  \end{section}

\begin{section}{Optical lattice Hamiltonian}\label{Hamiltonian}
Our starting point is the second-quantized Hamiltonian for bosonic atoms in an external trapping potential $H = H_0+H_{\rm int}$, where
\begin{align}
   H_0 &= \int d^3x \, \psi^\dagger(\mathbf{x}) \left(\frac{-\hbar^2}{2m_a}\nabla^2 + V_{op}(x,y)+  V_z(z)\right)\psi(\mathbf x)\nonumber         
\end{align}
and $H_{\rm int}=(g/2)\int d^3x\,\psi^\dagger(\mathbf x)\psi^\dagger(\mathbf x)\psi(\mathbf x)\psi(\mathbf x),$ where $\psi(\mathbf x)$ is the field operator at position $\bf x$, $m_a$ is the atomic mass and $\hbar$ is the reduced Planck's constant. The optical-lattice  potential in the $x$-$y$ plane is given by  $V_{op}(x,y) = -V_0\left( \cos^2k_Lx+\cos^2k_Ly+ 2\cos\theta\cos k_Lx\cos k_Ly\right),$
with $k_L$  the magnitude of the laser wave-vector and $\theta$  the phase difference between the counter propagating laser beams forming the optical lattice. The positive $V_0$ determines the lattice depth and $V_{op}(x,y)\leq 0$ for all $x$ and $y$.  
In addition, there is also a weak harmonic potential in the $z$ direction, $V_z(z) =  m_a\omega_z^2z^2/2$, with frequency $\omega_z$. The Hamiltonian $H_{\rm int}$ describes the atom-atom interaction where $g=4\pi\hbar^2a_s/m_a$, which is proportional to the s-wave scattering length $a_s$.  It is convenient to define the recoil energy $E_R=\hbar^2k_L^2/(2m_a)$. We study near degeneracies between the various excited bands. They occur when $V_0$ is between $5E_R$ and $10E_R$. A weak harmonic trap corresponds to $\hbar\omega_z\ll E_R$.
 
 The optical lattice potential $V_{op}(x,y)$ gives rise to a double-well optical-lattice, with adjacent wells of unequal well depth, depending on the value of $\theta$.
 Figure \ref{2D_lattice-a} shows an example of the 2D double well optical lattice formed by $V_{op}$. The two primitive lattice vectors are $\mathbf a_1  = a\left(\mathbf{\hat x}+\mathbf{\hat y} \right)$ and $\mathbf a_2  = a\left(\mathbf{\hat x}-\mathbf{\hat y} \right)$, where $a=\pi/k_L$ is the distance between neighboring shallow and deep wells. The reciprocal lattice vectors are $\mathbf{G}=\mathbf G_{m,n}=m\mathbf b_1+n\mathbf b_2$, with integers $m,n$, and primitive reciprocal lattice vectors $\mathbf{b}_1=k_L(\hat{\mathbf{x}}+\hat{\mathbf{y}})$ and $\mathbf{b}_2=k_L(\hat{\mathbf{x}}-\hat{\mathbf{y}})$. Figure \ref{kq-b} shows the first Brillouin zone as a function of quasi momentum $\mathbf K=(K_x,K_y)$. For convenience we define the quasi momenta $q_1=K_x+K_y$ and $q_2=K_x-K_y$. In this basis the first Brillouin zone is the square region $-\pi/a<q_1,q_2\leq\pi/a$.
    \begin{figure}
       \centering
       \subfloat[Part 1][]{\includegraphics[width=1.45in]{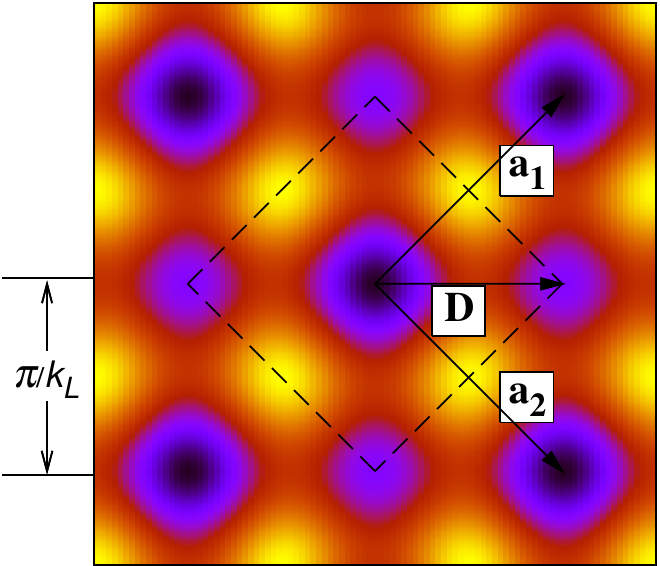}\label{2D_lattice-a}} 
       \hspace{5mm}
       \subfloat[Part 2][]{\includegraphics[width=1.52in,height=1.25in]{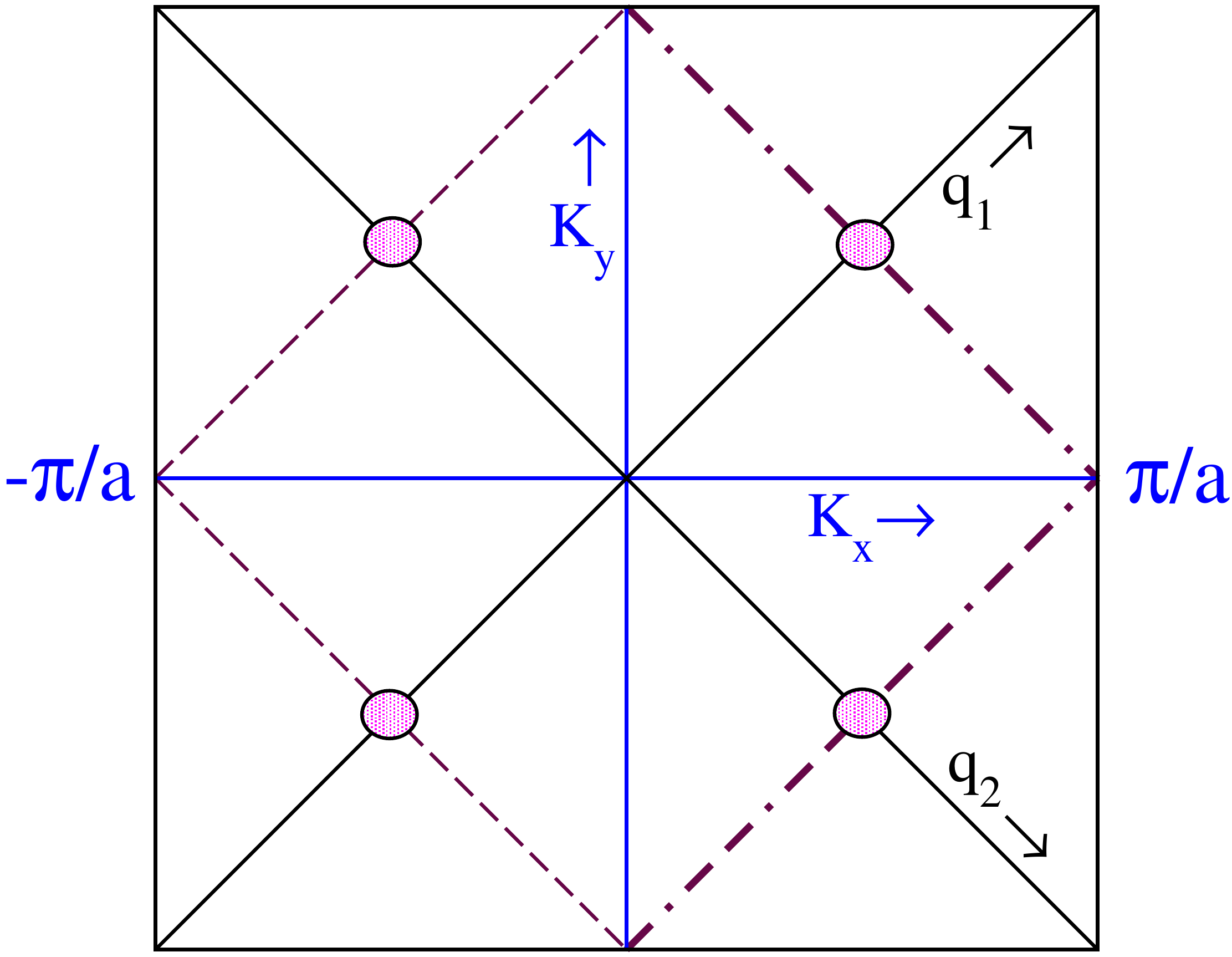}\label{kq-b}}
       \caption{(Color online) Panel(a) shows the contour graph of the 2D double-well optical lattice at $V_0=6E_R$ and $\theta=\pi/2.2$. The darker and brighter colors represent lower and higher energies, respectively. An alternating array of deeper and shallower wells are clearly visible. The origin of our coordinate system is at the central deep well. The primitive lattice vectors are $\mathbf{a}_1$ and $\mathbf{a}_2$ and the vector ${\bf D}=a\hat{\bf x}$ connects neighboring deep and shallow wells.The dashed lines enclose a unit cell. Panel(b) shows the first two Brillouin zones of the optical lattice as a function of quasi momentum $(K_x,K_y)$. The area inside the dashed region including the dashed-dotted edge represents the first zone. Moreover we define  $q_1=K_x+K_y$ and $q_2=K_x-K_y$. The circles represent the atoms at the four non-zero quasi momenta, where the second band has the minimum of energy.}
       \label{lattice}
     \end{figure}
  \end{section}

\begin{section}{Band structure in the tight binding model}\label{TB-model}
     An approximate band structure for the 2D single particle Hamiltonian,   $H_{op}=-(\hbar^2/2m_a)(\partial_x^2+\partial_y^2)+V_{op}(x,y)$ in the $xy$  plane is found from a tight binding model using localized harmonic oscillator wave functions as basis functions \cite{Altland}. We first expand $V_{op}(x,y)$ to second order in the coordinates around the minima of the deep and shallow well. To this order, the wells are isotropic and have harmonic oscillator energy $\hbar\omega_{d,s}  =\sqrt{4 V_0 E_R(1\pm \cos\theta)}$
 and harmonic length $l_{d,s}=\sqrt{\hbar/{m_a\omega_{d,s}}}$.
Here, the subscripts $d$ and $s$ stand for deep and shallow well, respectively.
 For $\theta<\pi/2$, an example of which is shown in Fig.~\ref{2D_lattice-a}, the deep well is located at the origin $(x,y)=(0,0)$ and the shallow well is located at $\mathbf{D}=a\hat{\mathbf{x}}$. 

\begin{figure}
  \centering
  \includegraphics[width=3.3in]{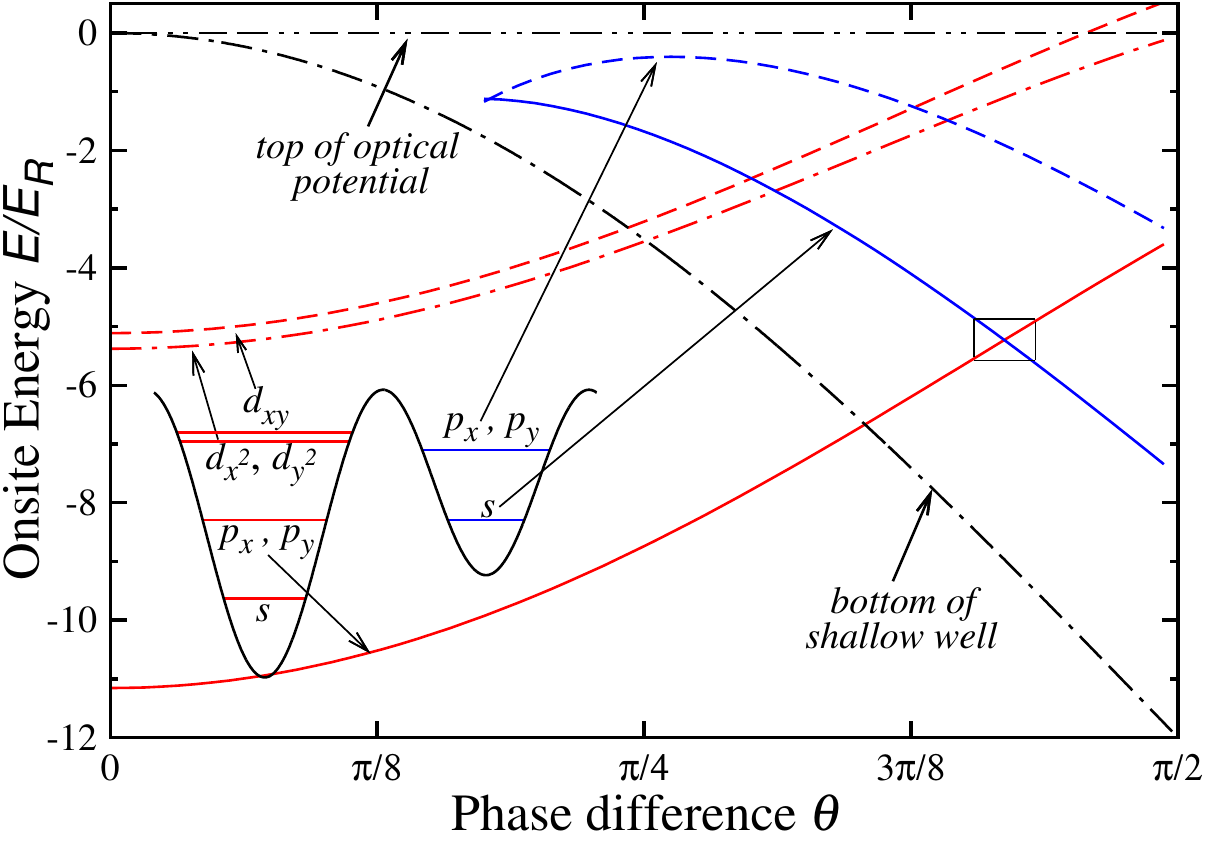}
  \caption{(Color online) Onsite energy in each well from tight binding model in units of $E_R$ as a function of the phase difference $\theta$ at $V_0=6E_R$. The energies are estimated in the harmonic oscillator approximation. The inset is a one-dimensional schematic of the asymmetric double well lattice, where horizontal lines in the wells represent the onsite energy levels. The various arrows map these  lines to the corresponding onsite energies. The zero of energy is at the maximum energy of the optical potential. For the deep well, $s$, $p$ and $d$ orbitals exist, except for $\theta\approx\pi/2$ where the $d$ orbital no longer exist.  The $s$ orbital in the deep well has an energy below $-12E_R$. The dashed line indicates the bottom of the shallow well. This  well does not support bound states for $\theta<0.18\pi$. The box denotes the point of degeneracy, $\theta=\theta_{res}\approx 0.42\pi$, between the $s$ state in the shallow well and the $p$ state in the adjacent deep well.}
  \label{onsite_energy}
  \end{figure}

 The harmonic oscillator wave functions for the deep and shallow  wells based on frequencies $\omega_d$ and $\omega_s$ are the local wave functions in our tight binding model. We choose $\phi_j(\mathbf r)$ to represent wave functions in the shallow well and $\chi_j(\mathbf r)$ for wave functions in the deep well. The subscript $j=s,p_x,p_y,d_{x^2},d_{y^2},d_{xy},\ldots$, where $s$ stands for the ground state and $p_x$ and $p_y$ represent the first two degenerate excited states with excitations along the $x$ and $y$ directions, respectively. The states $d_{x^2},d_{y^2},d_{xy}$ represent the doubly-excited $d$ orbitals, with two excitations along the $x$ direction, two excitation along the $y$ direction and one excitation in each direction, respectively. For lattice depths relevant in this paper, orbitals with higher excitations are not important.

 In the tight binding model, the local onsite energies $\langle\psi_j\vert H_{op}\vert\psi_j\rangle$ and $\langle\chi_j\vert H_{op}\vert\chi_j\rangle$ are expectation values of the total optical-lattice Hamiltonian, an improvement over solely using the harmonic oscillator approximation. A schematic of the local onsite energies in these wells is shown in the inset of Fig.~\ref{onsite_energy}. The onsite energy for the $p_x$ or $p_y$ state in the deep well is 
\begin{align}
  \label{eq:onsite_energy_deep}
  E_p & = \hbar\omega_d-V_0\left\{1+\left(1-k_L^2l_d^2\right)e^{-k_L^2l_d^2}\right.\nonumber\\
      &\qquad\qquad\qquad\,\,\,\, \left. +\, 2\cos\theta\left(1-\frac{k_L^2l_d^2}{2}e^{-k_L^2l_d^2/2}\right)\right\},
 \end{align}
and onsite energy of the $s-$orbital in the shallow well is 
\begin{align}
  \label{eq:onsite_energy_shallow}
  E_s & =\hbar\omega_s/2 -V_0\left\{1+e^{-k_L^2l_s^2}-2e^{-k_L^2l_s^2/2}\cos\theta\right\}.
\end{align}
  We are interested in parameter regimes $(V_0,\theta)$ where $E_p$ and $E_s$ are degenerate or nearly degenerate. 

Figure~\ref{onsite_energy} shows five onsite energies as a function of $\theta$ for $V_0=6E_R$.  Only energies close to $E_s$ and $E_p$ are shown. For this lattice depth and for $0<\theta<\pi/2$, the degeneracy occurs at $\theta=\theta_{res}\approx 0.42\pi$. For other values of $V_0$, the degeneracy shifts to a different $\theta$. We also note that the onsite energy of the $d_{xy}$ state is larger than that of the $d_{x^2}$ or $d_{y^2}$ state. 

From these local wave functions, we can construct extended  basis functions of definite quasi-momentum $\mathbf K$, $\langle\mathbf x\vert j, \mathbf K \rangle_s$ and $\langle\mathbf x\vert j, \mathbf K \rangle_d$ \cite{Altland}, which involves a sum over the total number of unit cells $M$. Here,
$\mathbf K$ is restricted to the first Brillouin zone, and the subscripts $s$ and $d$ on the kets stand for shallow and deep wells, respectively. The Hamiltonian $H_{op}$  conserves $\mathbf K$.

Close to the degeneracy point shown in Fig.~\ref{onsite_energy}, we can write down a tight binding Hamiltonian only taking into account the three basis functions $\vert s, \mathbf K \rangle_s$, $\vert p_x, \mathbf K \rangle_d$ and $\vert  p_y, \mathbf K \rangle_d$. The Hamiltonian is 
    \begin{equation}
\begin{pmatrix}\label{eq:TBHamiltonian}
        E_s & -2it\sin K_xa &-2it \sin K_ya\\
        2it\sin K_xa & E_p & 0\\
        2it\sin K_ya &0 & E_p
      \end{pmatrix}
\begin{array}{l}
   \vert s, \mathbf K \rangle_s\\\vert p_x, \mathbf K \rangle_d\\\vert  p_y, \mathbf K \rangle_d,
\end{array}         
    \end{equation}
\begin{figure}
  \centering
  \includegraphics[width=3.3in]{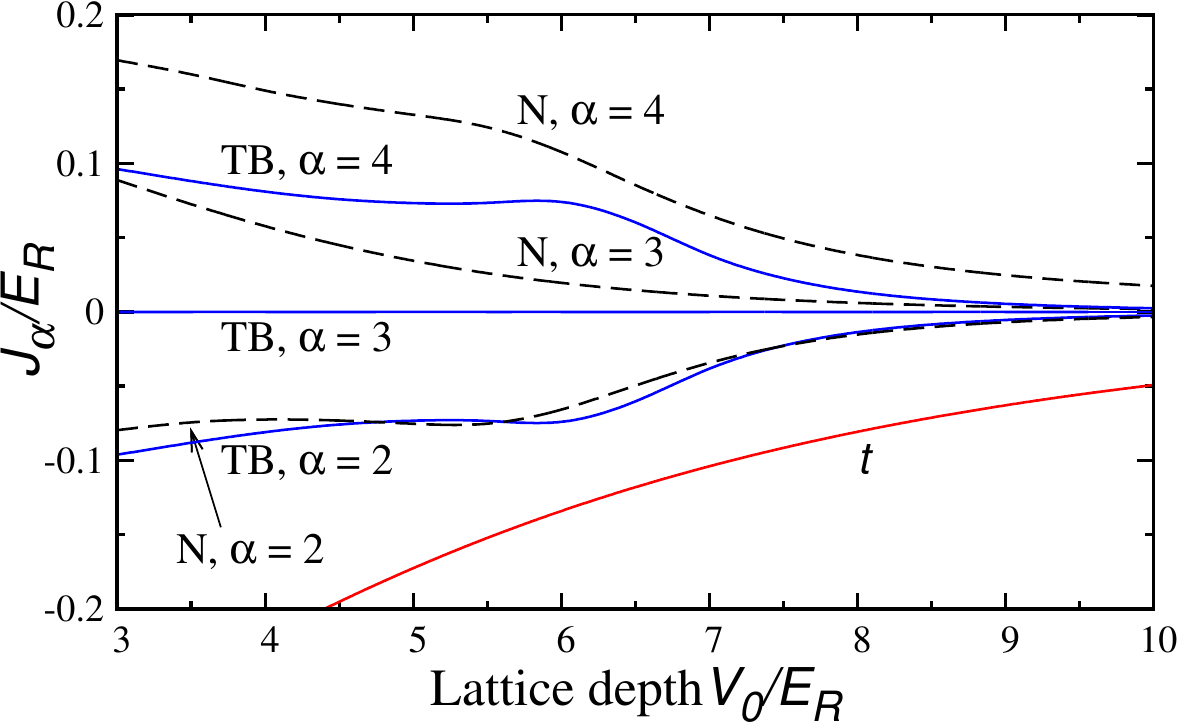} 
  \caption{(Color online) Tunneling energies $J_{\alpha}$ and $t$ as a function of lattice depth $V_0$ at $\theta=0.42\pi$. The solid blue and dashed black lines correspond to $J_{\alpha}$  between neighboring unit cells along the $x$ axis for $\alpha=2,3, 4$ based on tight binding (TB) and exact numerical (N) plane wave simulations, respectively. The tunneling for $\alpha=1$ is negligible on the scale of this figure. The solid red line shows the tunneling energy $t$ between the two wells in the unit cell.}
  \label{tunneling}
\end{figure}
with tunneling energy $t=\int d\mathbf{r}\varphi_s(\mathbf r-\mathbf{D}) H_{op}\chi_{px}(\mathbf r)$ between the adjacent wells. For $V_0$ between $5E_R$ and $10E_R$, the tunneling $t$ between neighboring deep and shallow wells is less than $E_R$. Simultaneously, we will require $t$ to be equal to or larger than $\hbar\omega_z$. The eigen energies of Eq.~\eqref{eq:TBHamiltonian} are
  $\epsilon_{2,4}(\mathbf K) =(E_s+E_p\mp\sqrt{\Delta^2+4b(\mathbf K)^2})/2$
 and $\epsilon_3(\mathbf K) = E_p$, where detuning $\Delta = E_s-E_p$ and  $b(\mathbf K)=2t(\sin^2K_xa+\sin^2K_ya)^{1/2}$ is quasi-momentum dependent. The corresponding Bloch or eigen functions are $\Phi_{\alpha\mathbf K}(\mathbf x)=\langle\mathbf x\vert\alpha,\mathbf K\rangle$, where the label $\alpha=2,3$ and $4$ stands for $2^{\it nd},3^{\it rd}$ and $4^{\it th}$ bands, respectively. The first band $\alpha=1$ has Bloch functions constructed from the local ground state in the deep well. The Bloch functions for the second band are $\vert 2,{\mathbf K} \rangle  = i\cos\theta_{\mathbf K} \vert  s, \mathbf K \rangle_s + \sin\theta_{\mathbf K}\vert +, \mathbf K \rangle_d$, where $\tan\theta_{\mathbf K} =2b(\mathbf K)/(-\Delta+\sqrt{\Delta^2+4b(\mathbf K)^2})$ and
\begin{align}\label{eigen-vector1}
\quad \vert + , \mathbf K\rangle_d&=\frac{\sin K_xa\vert  p_x, \mathbf K \rangle_d +\sin K_ya\vert  p_y, \mathbf K\rangle_d}{\sqrt{\sin^2 K_xa+ \sin^2 K_ya}}.
\end{align}

   \end{section}

   \begin{section}{Band structure using plane wave basis}\label{numerical-BS}
\begin{figure}
\subfloat{\includegraphics[width=3.3in]{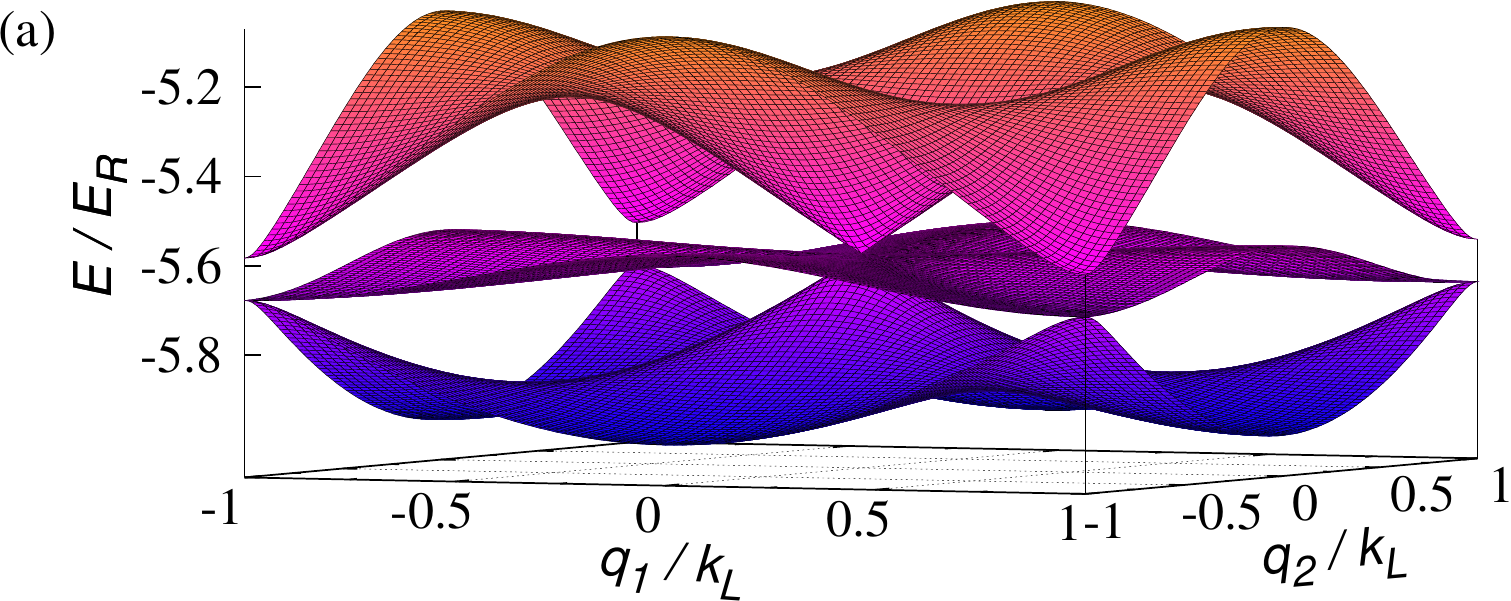} \label{band_hop-a}}\\
\subfloat{\includegraphics[width=3.3in]{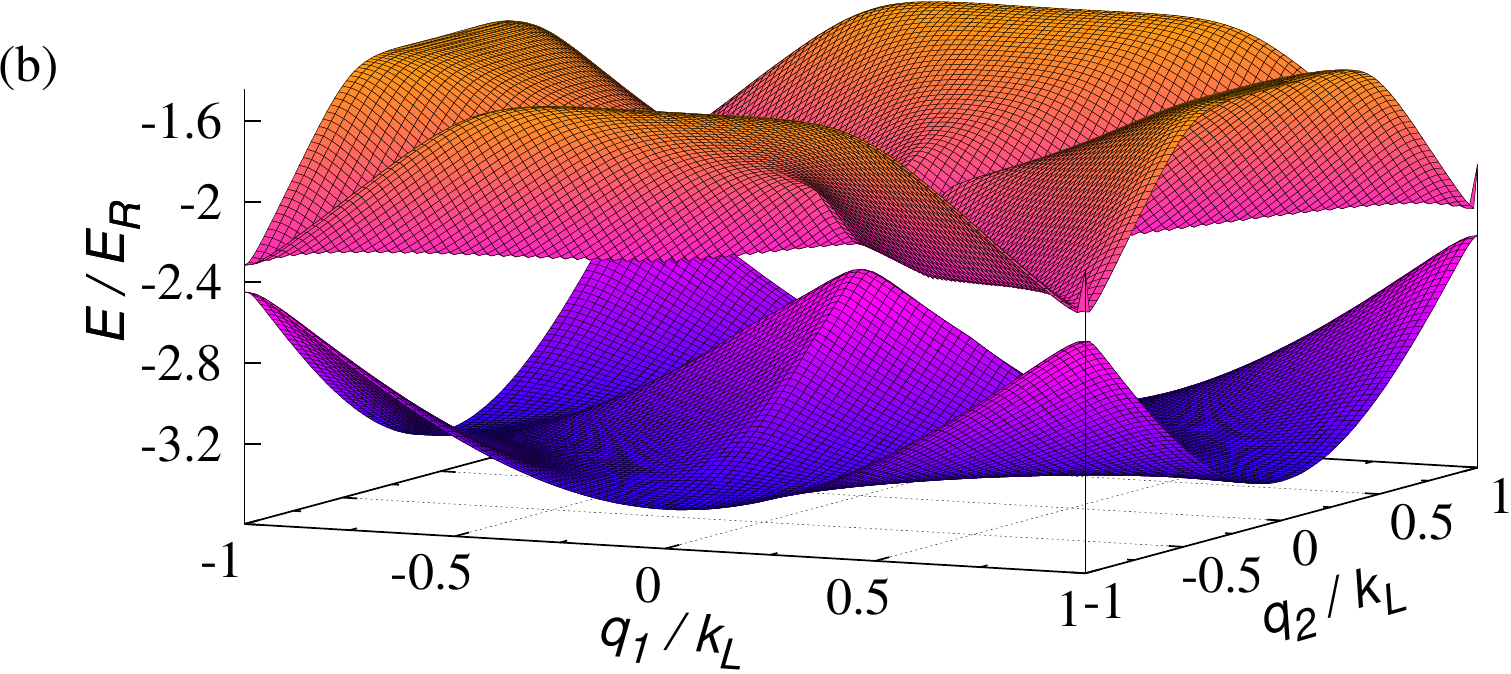} \label{band_hop-b}}
\caption{(Color online) Band structure and hopping parameters. Panel(a) shows the numerical band structure at $\vert V_0\vert=6E_R$ and $\theta=0.42\pi$ for the $2^{\it nd}$, $3^{\it rd}$ and $4^{\it th}$ bands as a function of the quasi momenta $\mathbf q=(q_1,q_2)$ in the first Brillouin zone. The second band has four degenerate minima at $\left(\pm k_L/2,\pm k_L/2\right)$, at the edge of the Brillouin zone. For the above set of parameters, the three bands touch at $\mathbf{K}=0$. Panel(b) shows the numerical band structure for the $5^{\it th}$ and $7^{\it th}$ bands, which provide decay channels for collision-aided decay of the condensate.}
\label{fig:fourfigures}
\end{figure}

     The 2D band structure for the single particle Hamiltonian $H_{op}$ can be numerically solved in the plane wave basis $\vert\mathbf k\rangle$, where the non-zero matrix elements between $\vert\mathbf K + \mathbf G_{m,n}\rangle$ and $\vert\mathbf K+\mathbf G_{m',n'}\rangle$ are $E_R\left\{\left(K_x/k_L+m  +n\right)^2+\left(K_y/k_L+m-n\right)^2\right\}-V_0$ for $m,n=m',n'$, $-V_0/4$ for $m=m'\pm1$, $n=n'\pm1$, and $-(V_0/2)\cos\theta$ for either of $m=m'\pm 1, n=n'$ or $m=m',n=n'\pm 1$.
It has energies $\epsilon_{\alpha}(\mathbf K)$ and Bloch functions $\Phi_{\alpha\mathbf K}(\mathbf x)=\sum_{\mathbf G}e^{i\mathbf{(K+G).r}}C_{\mathbf{G,K}}^{\alpha}$. These energies $\epsilon_{\alpha}(\mathbf K)$ for the $2^{\it nd}$, $3^{\it rd}$ and $4^{\it th}$ band are shown in Fig.~\ref{band_hop-a} for parameter set $V_0$ and $\theta$ where these three bands are degenerate at $\mathbf K=0$. 
The  $2^{\it nd}$ band has a maximum at $\mathbf K=0$ and has band minima at $\left(\pm k_L/2,\pm k_L/2\right)$, the points where a Bose-Einstein condensate forms, also shown in Fig.~\ref{kq-b}. Unlike in the tight binding model, the energy of the third band has a small momentum dependence. The band width of the $4^{\it th}$ band is significantly larger than what is predicted by the tight binding model.  Figure~\ref{band_hop-b} shows the higher $5^{\it th}$ and $7^{\it th}$ bands, which will provide collision-aided decay channels for the condensate formed in the second band. We have omitted other bands for clarity.

We ascertain the validity of the tight binding approximation by a comparison of the tunneling energy between unit cells with that obtained from the numerical band structure calculation. In particular, we compare the tunneling energy for band $\alpha$ between neighboring unit cells along the $x$ axis, given by $J_{\alpha} = -(1/M)\sum_{\mathbf{K}}\epsilon_{\alpha}(\mathbf K)e^{i2a\mathbf{K}\cdot\hat{\mathbf{x}}}$ and the sum is over the quasi-momenta $\mathbf K$ in the first Brillouin zone.  Results for bands $\alpha=2,3$ and $4$ and $\theta=0.42\pi$ are shown in Fig.~\ref{tunneling} over the relevant range of lattice depths where the bands can touch. We find that the two methods agree to within $10\%$ for the $2^{\it nd}$ band at lattice depth of $V_0=6E_R$ which is the point of degeneracy between the $s$ and $p$ states in adjacent wells. For other $\theta$ close to $0.42\pi$, the results are similar. For the two higher bands the agreement is less satisfactory as we have restricted ourselves to a tight binding model comprising of three localized states only.  
As we will show later, for the main results of this paper, the tight binding model is nevertheless sufficient. For example, the contribution of the $3^{\it rd}$ and $4^{\it th}$ bands to the thermodynamics of the problem will be small, and they do not provide channels for the condensate decay from the $2^{\it nd}$ band. Finally, we find that for larger well depths, the tunneling energy from the tight binding model is always smaller than the numerical estimate. This is because the dominant contribution to the hopping energy occurs in the classically forbidden region where the wave functions drop off exponentially. In these regions between the lattice sites, the anharmonic corrections to the potential, correctly treated in the numerical model, decrease the drop-off of the wave functions  and give rise to larger tunneling energies.

   \end{section}

\begin{section}{Thermodynamics for the non-interacting Bose gas}\label{Thermodynamics}
\begin{figure}
\centering
\subfloat{\includegraphics[width=3.3in]{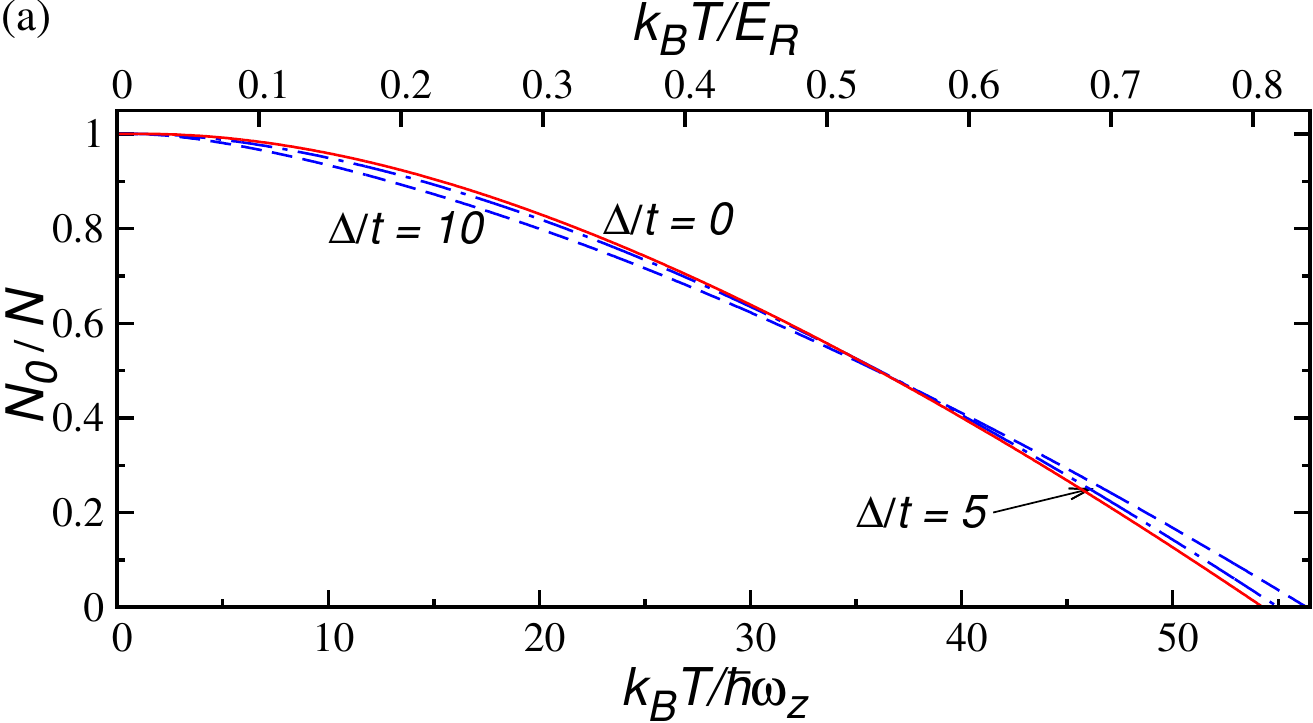} \label{fig:CondensatevsDelta}}\\
\subfloat{\includegraphics[width=3.3in]{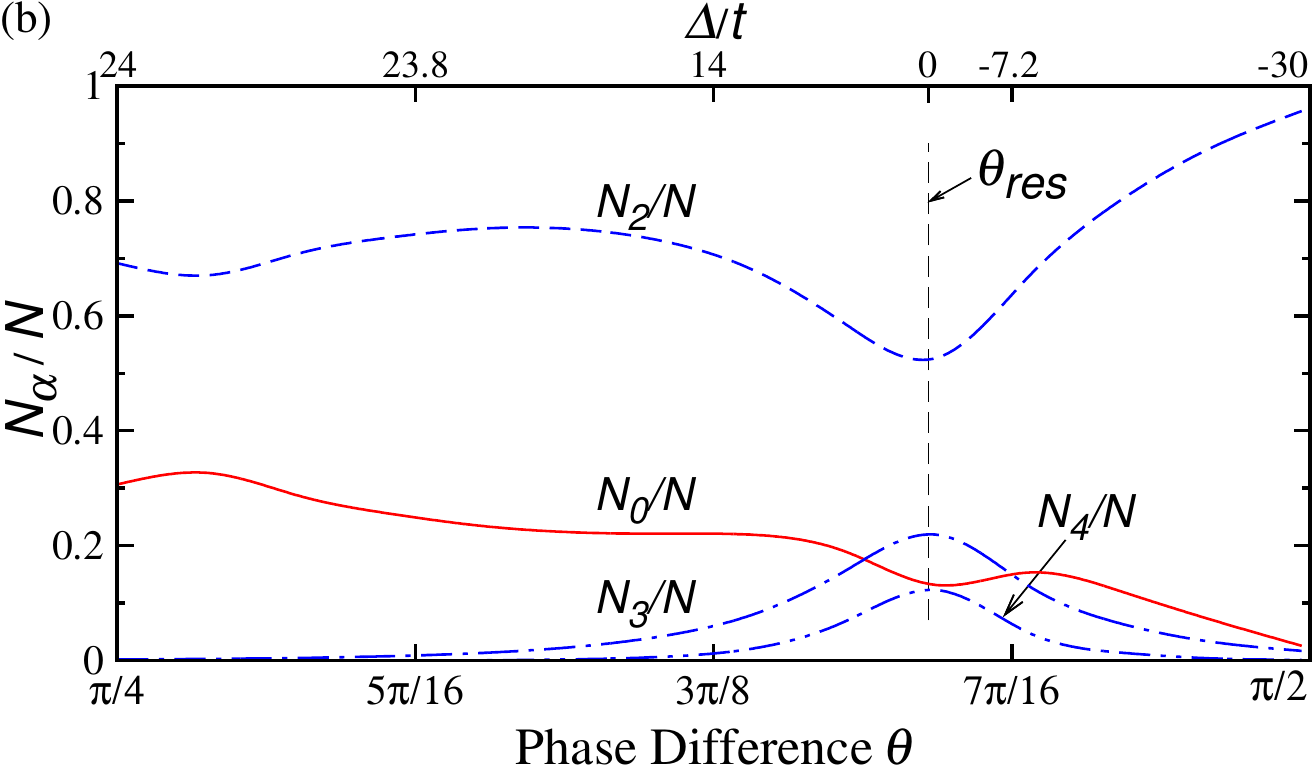} \label{fig:N0vsDelta}}
\caption{(Color online) Panel (a) shows the condensate fraction as a function of $k_BT/\hbar\omega_z$ (bottom axis) and $k_BT/E_R$ (top axis), for three values of $\Delta/t$. We used $V_0=6E_R$ and the values of $\Delta/t$ correspond to $\theta=0.420\pi,0.405\pi$, and $0.389\pi$ for increasing $\Delta$. The frequency of the harmonic trap along the $z$ direction is $30$Hz or $\hbar\omega_z=0.015 E_R$, which is a typical experimental value, leading to $t=10\hbar\omega_z$ for $\Delta=0$. Finally, $\nu=N/M=200$.  Panel (b) shows $N_0/N$ and the non-condensed fractions ${\cal N}_{\alpha}/N$ as a function of $\theta$ and $\Delta/t$ for $k_BT= 50\hbar\omega_z=0.74E_R$. The non-linear $x$ axis on the top of the figure shows the corresponding $\Delta/t$ values. Between $\theta=\pi/4$ and $5\pi/16$ the value of $\Delta/t$ do not change significantly. Other parameters are as in Panel (a). The value of $\vert\Delta\vert/t$ is zero at the point of degeneracy with $\theta=\theta_{res}$, and increases as we move away from resonance.}
\label{fig:fourfigures} 
\end{figure}

We discuss the formation of a Bose-Einstein condensate for non-interacting bosons in the first excited or second band of the optical lattice. We study temperatures $T$ such that $k_BT> t\gg\hbar\omega_z$, but $k_BT$ is much less than the large band gap between the $4^{\it th}$ and $5^{\it th}$ band. Thus, for our parameters,  we only need to consider population in the $2^{\it nd}$, $3^{\it rd}$ and $4^{\it th}$ bands. As we will show, the fraction of condensed atoms is then small and therefore the picture of non-interacting bosons is appropriate. Furthermore, we will only present results using the tight binding model. This is sufficient as it  gives a good representation of the second band and most atoms are in this band. Also, with realistic approximations the model leads to analytic expressions which allows us to study the parameter dependencies in detail. 

Following Ref.~\cite{castin_bose-einstein_2001}, the total atom number $N=N_0+\sum_{\alpha=2,3,4}{\cal N}_{\alpha}$, where $N_0$ is the number of atoms in the condensate at the minima of the second band, and 
\begin{align}
  \label{eq:N'2nd}
  {\cal N}_{\alpha}&=2M\frac{a^2}{(2\pi)^2 }\sum_{k=1}^\infty \sum_{n=0}^\infty z^k \int \frac{\ud^2 \mathbf{q}}{2}e^{-k\beta (\epsilon_{\alpha}(\mathbf q) +n\hbar\omega_z)}
\end{align}
is the number of (non-condensed) atoms in band $\alpha$ excluding the condensed atoms. Here $z$ is the fugacity, satisfying $0\leq z\leq 1$ and $\beta=1/k_BT$. The sum $n$ is over all harmonic oscillator levels in the $z$ direction and the integral over quasi momentum $\mathbf K$ or equivalently $\mathbf q$ is over the first Brillouin zone. The sum over $k$ originates from a series expansion of the Bose distribution. We have set the zero of energy of the dispersion $\epsilon_{\alpha}(\mathbf q)=\epsilon_{\alpha}(\mathbf K)$ at the minimum of the second band. Similarly the zero of energy of the harmonic oscillator is at the $n=0$ level.

 The maximum number of non-condensed atoms ${\cal N}_{\alpha}$ occurs at $z=1$. Hence, condensation occurs when $N$ is larger than the sum of the maximum non-condensed atom number in each band and thus for the remainder of the paper we will assume $z=1$ in Eq.~\eqref{eq:N'2nd}.  The integral over $\mathbf q$ is straightforward  for the third band as its energy is independent of $\mathbf q$, but requires more thought for the other two bands. We are interested in the number of atoms occupying energy levels close to the bottom of the second band where $1+\cos q_1a\cos q_2a=0$. Taylor expanding the energy of both the $2^{\it nd}$ and $4^{\it th}$ bands to first order in $1+\cos q_1a\cos q_2a$ gives  $\epsilon_2(\mathbf q)\approx [t/F(\Delta/t)](1+\cos q_1a\cos q_2a))$ and $\epsilon_4(\mathbf q)\approx 4t\F-[t/F(\Delta/t)](1+\cos q_1a\cos q_2a))$, where $F(x)=\sqrt{2+x^2/16}$. 
The integrand $\exp(-k\beta\epsilon_{\alpha}(\mathbf q))$ is separated into a momentum dependent and an independent part. The momentum independent part becomes a simple pre-factor. Using that $\beta t/\F\ll 1$, the momentum dependent part of the integral is solved to $O([\beta t/\F]^2)$. We have checked that this approximation to the integral over $\mathbf q$ is comparable to a numerical estimate of the same. We can perform the sum over $n$ exactly. Then assuming that $\hbar\omega_z\ll k_BT$, the sum over $k$ can be performed.

Finally, we have
 \begin{align}\label{Nprime}
{\cal N}_2& \approx M\frac{k_BT}{\hbar\omega_z}\log\left[\frac{1}{1-e^{-\frac{\beta t}{\F}}}\right],\\
  {\cal N}_{3}&\approx M\frac{k_BT}{\hbar\omega}\log\left[\frac{1}{1-e^{-2\beta t\F}}\right],\\
{\cal N}_4&\approx M\frac{k_BT}{\hbar\omega}\log\left[\frac{1}{1-e^{-\beta t\left(4\F-\frac{1}{\F}\right)}}\right],
\end{align}
which are even functions of $\Delta$.

Figure~\ref{fig:CondensatevsDelta} shows the dependence of the condensate fraction $N_0/N$ on the temperature for various values of detuning $\Delta/t$, where the $s$ orbital in the shallow well is or is nearly degenerate with the $p$ orbitals in the deep well. For  a lattice depth $V_0$, the different ratios of $\Delta/t$ can be obtained by appropriately choosing $\theta$. The temperature at which $N_0/N\to 0$ is the critical temperature $T_c$. We see that $k_BT_c\gg\hbar\omega_z$ for reasonable $\Delta/t$ and, thus, is consistent with the approximations that we have made. Moreover, our assumption of non-interacting atoms is valid for $T$ near $T_c$ where the fraction of atoms in the condensate is much smaller than one. 

Figure~\ref{fig:N0vsDelta} shows the dependence of $N_0/N$ and ${\cal N}_{\alpha}/N$ on $\theta$ and $\Delta/t$ for $T$ just below $T_c$. We first note that for $\theta$ close to $\theta_{res}$, the dependence of $\Delta/t$ with $\theta$ is linear. This linearity persists for $\theta_{res}\le\theta\le\pi/2$, but for $\theta<\theta_{res}$, $\Delta/t$ increases and then saturates. For the parameters shown the number of atoms in the second band $N_0+{\cal N}_2$ is larger than $60$\% of all atoms. At the degeneracy point, $\Delta=0$ and $N_0+{\cal N}_2$ is smallest. At this point, the bands touch, and consequently the atom population in the $3^{\it rd}$ and $4^{\it th}$ bands are highest. Nevertheless, the populations in bands $\alpha=3$ and $4$ are sufficiently small such that the tight binding approximation remains valid.

\end{section}

\begin{section}{Lattice transformation and final temperature}\label{lattice transformation}
  \begin{figure}
    \centering 
    \includegraphics[width=3.3in]{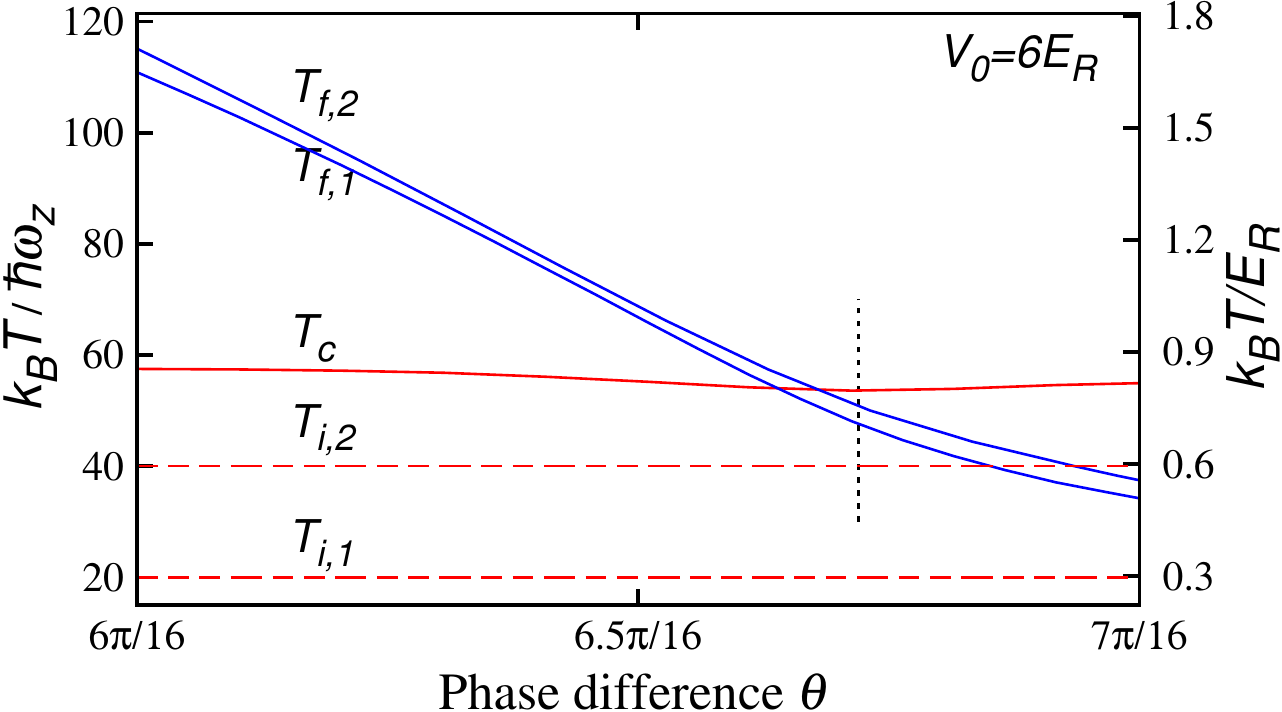}
    \caption{(Color online) The final temperature $T_{f,1}$ and $T_{f,2}$ (solid blue lines) in units of $\hbar\omega_z/k_B$ (left axis) or $E_R/k_B$ (right axis) after the lattice transformation as a function of the phase difference $\theta$ for two initial temperatures, $T_{i,1}=20\hbar\omega_z/k_B$ and $T_{i,2}=40\hbar\omega_z/k_B$, respectively. (Equivalently, we have $T_{i,1}=0.3E_R/k_B$ and $T_{i,2}=0.6E_R/k_B$, respectively). We use  atom number per lattice site $\nu=N/M=200$ and lattice depth $V_0=6E_R$. The initial temperatures are shown as dashed horizontal lines. The vertical dotted lines represent the angle $\theta$ at which the $s$ and $p$ orbitals of adjacent wells are degenerate. For fixed atom number the critical temperature $T_c$ (solid red line) of the final lattice weakly depends on the lattice parameters but is independent of $T_i$.}
    \label{fig:final-temperature}
  \end{figure}
As discussed in Section \ref{Introduction}, populating the higher bands is a two step process. In step $1$, the angle $\theta$ is chosen such  that the deep wells are significantly deeper than the neighboring shallow wells, and the atoms are confined to the local ground states of these deep wells, as shown in Fig.~\ref{step1-a}. The atoms are confined in the transverse directions, but distributed in the $z$ harmonic oscillator states with a temperature $T_i$. In step $2$, $\theta$ is changed to reach the final lattice configuration. We model this change to be fast with respect to tunneling energies between adjacent wells, but adiabatic with respect to the onsite energies of the well to which the atoms have been confined. Thus, just after the final $\theta$ is reached, the atoms are still confined to the ground $s$ state of these wells, which now happen to be the shallow wells (Fig.~\ref{step1-a}). We refer to this state as the initial state just after the lattice transformation, and the temperature for this state is still $T_i$. With the $s$ orbitals now tuned to resonance with the adjacent $p$ orbitals, the atoms gets distributed to the entire $2$D lattice as well as to harmonic levels in the transverse direction due to tunneling and elastic collisions. The atoms thermalize, and condenses to the four quasi-momenta at the edges of the first Brillouin zone of the second band. We refer to this state as the final state, and the temperature reached after thermalization as $T_f$.

We have calculated the final temperature using conservation of total number of atoms and total energy, and find an analytic expression between $T_i$ and $T_f$. Figure~\ref{fig:final-temperature} shows the dimensionless final temperature $k_BT_f/\hbar\omega_z$ as a function of the phase difference $\theta$. We see that for the two initial temperatures shown in the figure the final temperature $T_f$ is a  monotonically decreasing function of $\theta$, can lie either above or below the critical temperature $T_c$, but is nearly always larger than $T_i$.  At the point of degeneracy between the adjacent $s$ and $p$ orbitals, $T_f$ lies just below the critical temperature $T_c$, and the gas remains condensed. Nevertheless, this temperature is larger than the corresponding initial temperature.  We have $T_f>T_i$ since the energy of the initial state $E_s$ is greater than the bottom of the second band, $\epsilon_2(\K)$ at quasi-momentum $\K$. Elastic collisions then redistribute this excess energy over the harmonic oscillator levels as well as the different quasi momenta in the $2D$ plane and the temperature will increase.
On the other hand the energy difference $E_s-\epsilon_2(\K)$ decreases with $\theta$ so that $T_f$ decreases with $\theta$.

Thus the lattice transformation typically leads to heating of the atom cloud. For a given atom number $\nu$ and initial temperature $T_i$, there is a limited range of final lattice parameters $V_0$ and $\theta$ for which the final temperature $T_f$ lies below the corresponding critical temperature $T_c$. At the point of degeneracy of the adjacent $s$ and $p$ orbitals, even for significantly low initial temperatures $T_i$, the final temperature $T_f$  is large and lies close to the corresponding critical temperature $T_c$. At this final temperature, the fraction of condensed atoms will be small, and thus our picture of non-interacting atoms should hold.
\end{section}

\begin{section}{Lifetime estimate of the condensate}\label{lifetime}
\begin{figure}
    \centering
    \subfloat{\includegraphics[width=3.3in]{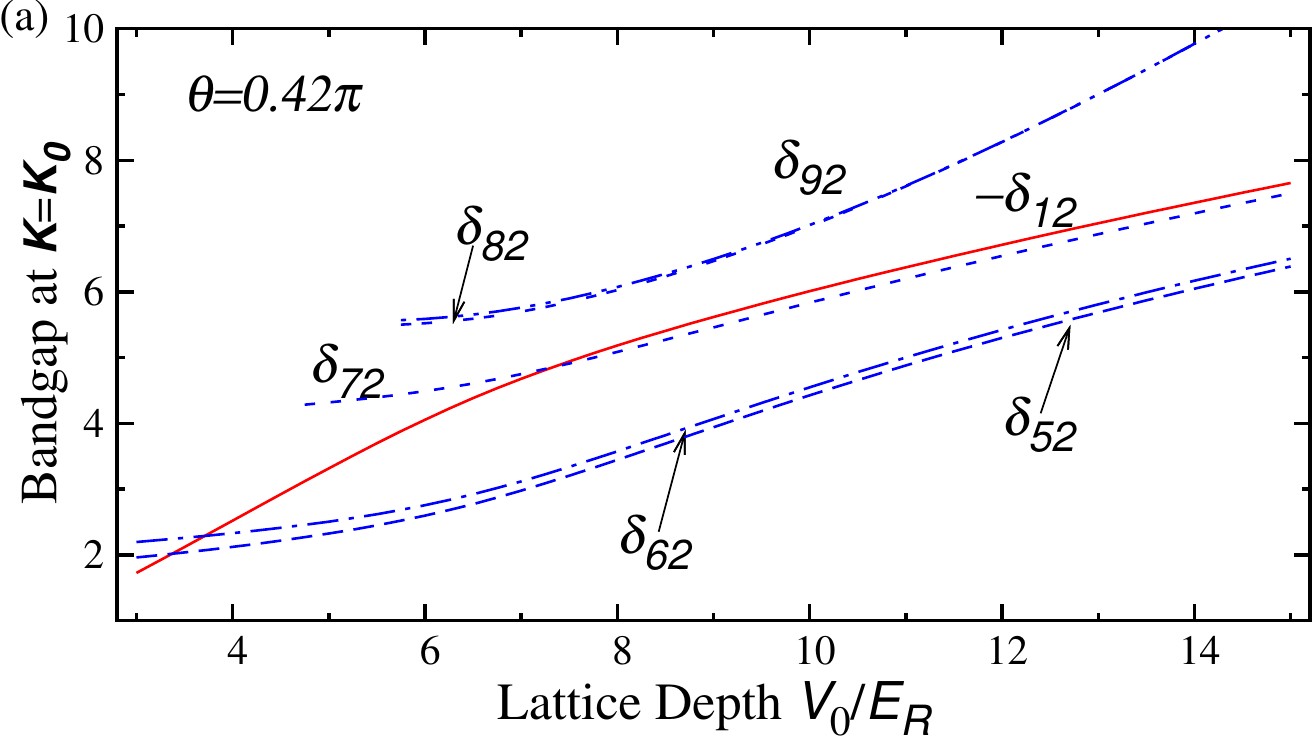}\label{bandgap}}\\
    \subfloat{\includegraphics[width=3.3in]{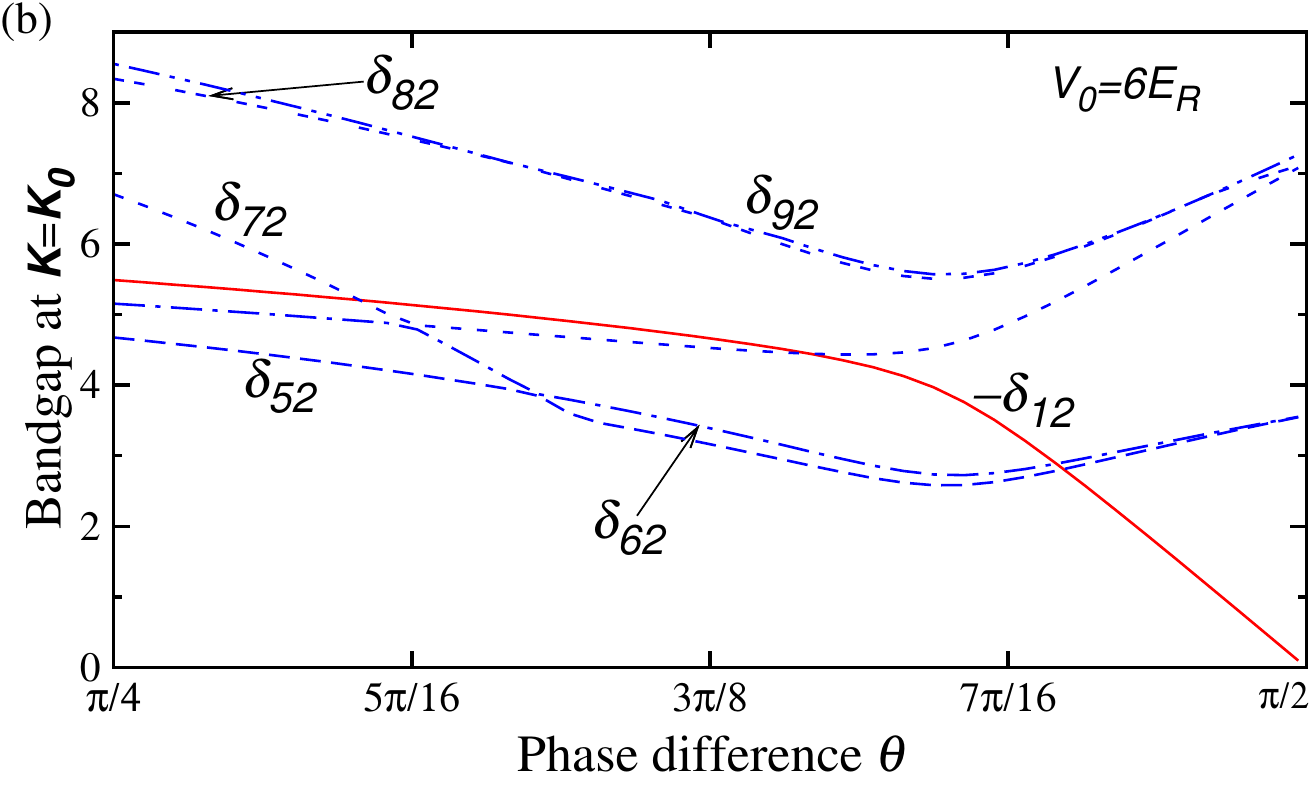}\label{bandgap-v6}}
    \caption{(Color online) Band gaps $\delta_{\alpha 2}=\epsilon_{\alpha}(\K)-\epsilon_{2}(\K)$ between band $\alpha$ and the $2^{\it nd}$ band  at one of the condensate momenta. Panel (a) show the band gaps as a function of lattice depth  at $\theta=0.42\pi$. For $\alpha=1$ (solid red line) we plot minus of the band gap. Energy conservation requires that allowed decay processes satisfy $\delta_{\alpha 2}\le-\delta_{12}$. In other words, for allowed processes, curves $\delta_{\alpha 2}$  with $\alpha>2$ must lie below the solid red line. The excess energy $\delta_{12}+\delta_{\alpha 2}$ is lost to the oscillator in $z$ direction. Bands $\alpha\geq 7$ only exist for sufficiently deep wells. The angle $\theta$ is chosen such that for $V_0=6E_R$, the $s$ and $p$ orbitals in adjacent wells are  degenerate. Panel (b) show the band gaps as a function of the phase difference $\theta$ at lattice depth $V_0=6E_R$. Depending on the value of $\theta$, there are one or more allowed decay processes.}
    \label{bandgaps}
  \end{figure}

The atoms in the higher bands decay to the ground band by atom-atom collisions. In this section we will determine the rate for these processes. To obtain the decay rates, we use the picture of a discrete level coupled to a continuum as discussed in Complement $\rm{C_I}$ in Ref.~\cite{Cohen_Tannoudji}. These rates can also be obtained using the Fermi Golden rule. First, we rewrite our interaction Hamiltonian $H_{\rm int}$ in terms of the eigenstates or modes of $H_0$. In other words, we expand  bosonic operators $\psi(\mathbf{x})=\sum_{\alpha{\mathbf{K}}m}\Phi_{\alpha\mathbf{K}}(\mathbf{x})\varphi_m(z)a_{\alpha\mathbf{K}m}^{}$, where $a_{\alpha\mathbf{K}m}^{}$ is the annihilation operator of an atom in Bloch function $\Phi_{\alpha\mathbf{K}}(\mathbf{x})$ and the $m^{\it th}$ harmonic oscillator wave function $\varphi_m(z)$  along the  $z$ direction.

Our initial state is a incoherent thermal mixture of normalized discrete states $\vert\Phi\rangle=(\sum_{\K} \eta_{\K} a_{2,\K,0}^{\dagger})^{N_0}(a_{2,\mathbf{K},m'}^{\dagger})^{n(\mathbf K,m')}\vert 0\rangle/Z$ where $\vert 0\rangle$ is the vacuum of no atoms and $Z=\sqrt{N_0!n(\mathbf{K},m')!}$. These states $\vert\Phi\rangle$  only contain atoms in the second band. Decay from atoms in other bands is less important as in Section V we showed that most atoms are in the second band. Atoms in modes $2,\K,0$ are the condensed atoms while atoms in modes $2,\mathbf K,m'$ with $\mathbf K,m'\neq \K,0$ are non-condensed atoms. The superposition $\sum_{\K} \eta_{\K} a_{2,\K,0}^{\dagger}$ is over the two quasi-momenta $\K=(k_L/2, k_L/2)$ and $\K=(k_L/2,-k_L/2)$, where the energy of the $2^{\it nd}$ band is minimal, and the coefficient $\eta_{\K}=1/\sqrt{2}$ when the momentum in the $x$ and $y$ direction has the same sign and is $i/\sqrt{2}$ otherwise \cite{wirth_evidence_2010},\cite{cai_complex_2011}. The positive integer $n(\mathbf K,m')$ is the number of atoms in quasi momentum $\mathbf K$ and harmonic oscillator level $m'$ given by the Bose distribution. 

Each of the normalized discrete states decays to the continuum states $\vert\mathbf Q\rangle$  given by $a_{1,\mathbf{K}_1,m}^{\dagger}a_{\alpha,\mathbf{K}_2,n}^{\dagger}(\sum_{\mathbf{K}_0}\eta_{\mathbf{K}_0}a_{2\K,0})^2\vert \Phi\rangle/\sqrt{N_0(N_0-1)}$ and $a_{1,\mathbf{K}_1,m}^{\dagger}a_{\alpha,\mathbf{K}_2,n}^{\dagger}a_{2\K,0} \,a_{2,\mathbf{K},m'}\vert \Phi\rangle/\sqrt{N_0n(\mathbf K,m')}$, which  corresponds to states where either two atoms or one atom decays from the condensate, respectively, to other allowed modes. Energy conservation requires that at least one of the atoms in the continuum state occupies the ground $\alpha=1$ band, while the second atom can either be in the ground or any higher band. Discrete states $\vert\Phi\rangle$ and $\vert\Phi'\rangle$ decay to different orthogonal continuum states $\vert\mathbf Q\rangle$ and $\vert\mathbf Q'\rangle$, respectively.

Certain decay processes involving matrix elements with bands $\alpha>2$ are only energetically allowed for sufficiently large lattice depths or large detuning $\Delta$ between adjacent $s$ and $p$ orbitals. In Fig.~\ref{bandgap} we plot the band gap $\delta_{\alpha 2}=\epsilon_{\alpha}(\K)-\epsilon_{2}(\K)$ as a function of the lattice depth $V_0$ for a certain angle $\theta$.  Due to energy conservation, decay is only allowed when $\delta_{12}\geq -\delta_{\alpha 2}$. For typical $\theta$ and $V_0$, decay processes involving bands $\alpha>7$ do not exist, and those involving the $7^{\it th}$ band are allowed only beyond a certain lattice depth. Decay process involving bands $\alpha=2,3$ and $4$ are ruled out by parity considerations, and hence not shown in Fig.~\ref{bandgap}. Using similar arguments, we can check which decay processes contribute at a particular $\theta$ from Fig.~\ref{bandgap-v6}, where we plot $\delta_{\alpha 2}$ as a function of phase difference $\theta$ for a certain lattice depth $V_0$.
    
The interaction  Hamiltonian takes the form
\begin{align}
\label{eq:Hint}
  H_{int}= &\,\frac{g}{2}\sum I_{\alpha\beta\alpha'\beta'}^{mnm'n'}(\mathbf{K}_1,\mathbf{K}_2,\mathbf{K}_3,\mathbf{K}_4)\nonumber\\
          &\,\times a_{\alpha\mathbf{K}_1m}^{\dagger}a_{\beta\mathbf{K}_2n}^{\dagger}a_{\alpha'\mathbf{K}_3m'}^{}a_{\beta'\mathbf{K}_4n'}^{},
\end{align}
where the sum is over all indices and quasi-momenta. We can write  $I_{\alpha\beta\alpha'\beta'}^{mnm'n'}(\mathbf{K}_1,\mathbf{K}_2,\mathbf{K}_3,\mathbf{K}_4)=O_{m'n'}^{mn}P_{\alpha'\beta'}^{\alpha\beta}(\mathbf{K}_1,\mathbf{K}_2,\mathbf{K}_3,\mathbf{K}_4)$, where $O_{m'n'}^{mn}$ is an integral over four harmonic-oscillator wave functions along the $z$ axis, and $P_{\alpha'\beta'}^{\alpha\beta}(\mathbf{K}_1,\mathbf{K}_2,\mathbf{K}_3,\mathbf{K}_4)$ involves an integral over four  Bloch functions.

 The function $O_{m'n'}^{mn}$ can be found exactly \cite{edwards_zero-temperature_1996} and is non-zero only when $n+m+m'+n'$ is even. For our calculations, we only require matrix elements with $n'=0$ and $0\leq m'\ll m,n$ and an approximate form of the result proves to be more useful  for summing up the contributions to the loss rate. For the square of the function, we find $(O_{00}^{mn})^2\approx e^{-s^2/r}/(2\pi^2 l_z^2r)$ for  $m'=0$, and
 \begin{equation}\label{Iharmonic-1}
   \left(O_{m'0}^{mn}\right)^2\approx
 \frac{1}{2\pi^2 l_z^2r\Gamma(m'+1)}\left(\frac{s^2}{r}\right)^{m'}e^{-s^2/r},
  \end{equation}
for $m'\geq 1$. Here $2r=m+n$ and $2s=m-n$. The selection rules for $O_{m'0}^{mn}$ require that $r$ and $s$ are integers when $m'$ is even and half-integers when $m'$ is odd. Figure~\ref{fig:harmonic} shows a comparison of the exact  and approximate form of the function $O_{m'0}^{mn}$ for a fixed large $r$. For all $m'$, the approximate form is in good quantitative agreement with the exact.  Furthermore, we see that 
$O_{m'0}^{mn}$ is significant only when $s\ll r$, i.e. when $n$ and $m$ are relatively close. The function peaks at $s=\pm\sqrt{m'r}$.
\begin{figure}
  \centering
  \includegraphics[width=3.3in]{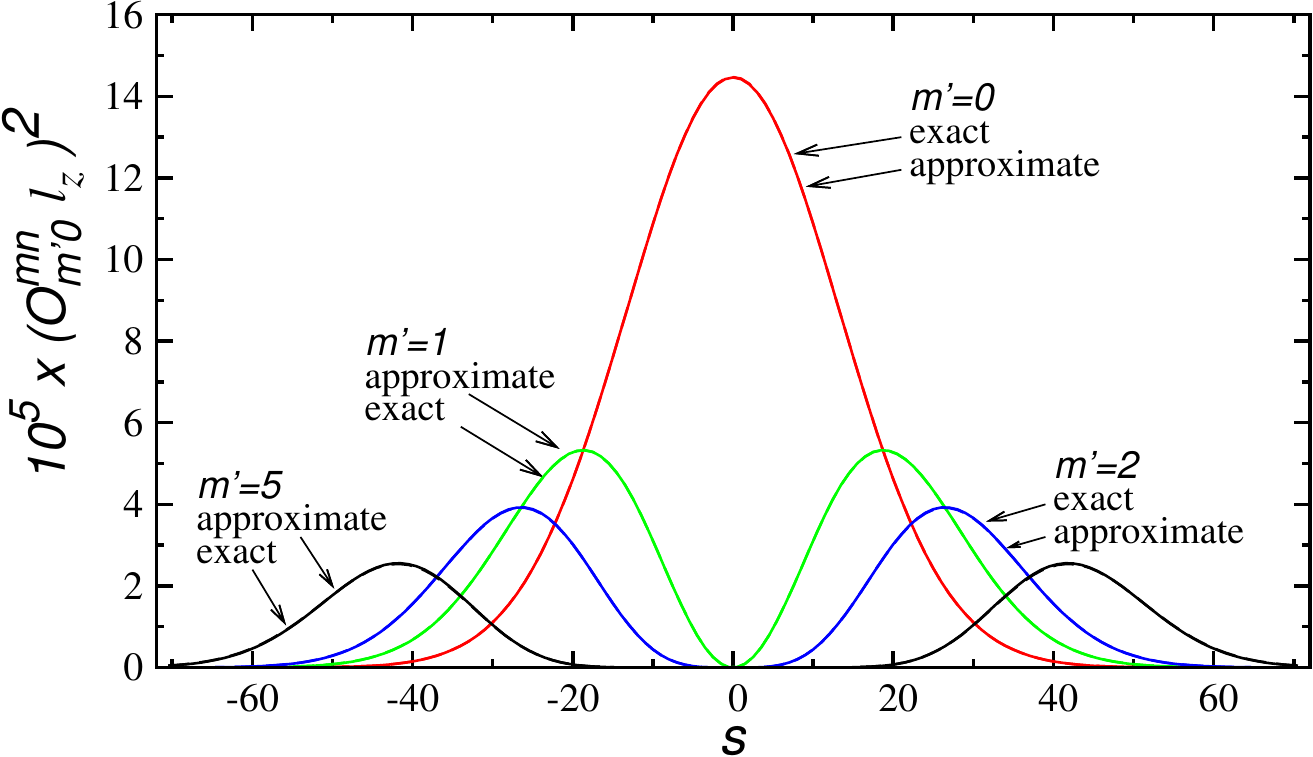}
  \caption{(Color Online) The square of the matrix element $O_{m'0}^{mn}$ as a function of $s=(m-n)/2$ with $r=(m+n)/2=350$. This large value for $r$ is typical for both atoms decaying from the second to the lowest band. The curves are for $m'=0,1,2$, and $5$. The solid curves are exact analytic evaluation of $O_{m'0}^{mn}$ while the indistinguishable  dashed curves correspond to approximate evaluations assuming $n+m\gg m'$.}
  \label{fig:harmonic}
\end{figure}

The function $P_{\alpha'\beta'}^{\alpha\beta}(\mathbf{K}_1,\mathbf{K}_2,\mathbf{K}_3,\mathbf{K}_4)$ is either numerically evaluated in the plane wave basis or  is analytically determined in the tight binding model. Bands $\alpha\le 4$ are always treated in the tight binding model while higher bands are treated numerically in the plane wave basis. The function $P_{\alpha'\beta'}^{\alpha\beta}$ is non-zero when $\mathbf{K}_1+\mathbf{K}_2=\mathbf{K}_3+\mathbf{K}_4\mod{\bf G}$. For our calculations, we only require matrix elements with $\alpha'=\beta'=2$ and one or both of $\mathbf K_3$, $\mathbf K_4$ equal the condensate momenta $\K$. We then find 
\begin{align}
  P_{22}^{\alpha\beta}(\mathbf{K}_1,\mathbf{K}_2,\mathbf{K},\mathbf{K}_0) & = \frac{\sin\theta_{\K}\sin\theta_{\mathbf K}}{4\sqrt 2M\pi l_d^2}f^{\alpha\beta}(\mathbf{K}_1,\mathbf{K}_2,\mathbf{K}).\nonumber     
\end{align}
 For $\alpha,\beta=1,1$ we find analytically $f^{11}(\mathbf{K}_1,\mathbf{K}_2,\mathbf{K})=(\sin K_xa\pm\sin K_ya)/\sqrt{\sin^2K_xa+\sin^2K_ya}$ and the $\pm$ sign corresponds to the case  $\K=(\pi/2a,\pi/2a)$ and $(\pi/2a,-\pi/2a)$, respectively. This matrix element is independent of $\mathbf{K}_1$ and $\mathbf{K}_2$ as $\alpha=1$ is a flat band in the tight binding approximation. If $\mathbf{K}_0=(\pi/2a,\pi/2a)$ and $\mathbf{K}=(\pi/2a,-\pi/2a)$, the matrix element is zero and vice-versa. Thus, collisions involving atoms in two inequivalent condensate quasi-momenta do not lead to  condensate decay.

The matrix elements $P_{22}^{1\alpha}$ with $\alpha=2,3,4$ are zero because of parity considerations, and this has also been checked numerically. For $P_{22}^{1\alpha}$ with $\alpha>4$, the integral is solved numerically using the tight binding wave functions for bands $1$ and $2$ but the numerical wave functions for band $\alpha$ expressed in terms of the  $C_{\mathbf G,\mathbf K}^{\alpha}$ defined in Sec.~\ref{numerical-BS}. We find that $P_{22}^{11}>P_{22}^{1\alpha}$ for $\alpha\neq 1$, independent of quasi-momenta $\mathbf{K}$.  This is also confirmed by the tight binding model. Finally, close to $\theta=\theta_{res}$, for bands $\alpha>4$, $P_{22}^{17}$ is the most dominant term. 

\begin{figure}
\centering
\subfloat{\includegraphics[width=3.3in]{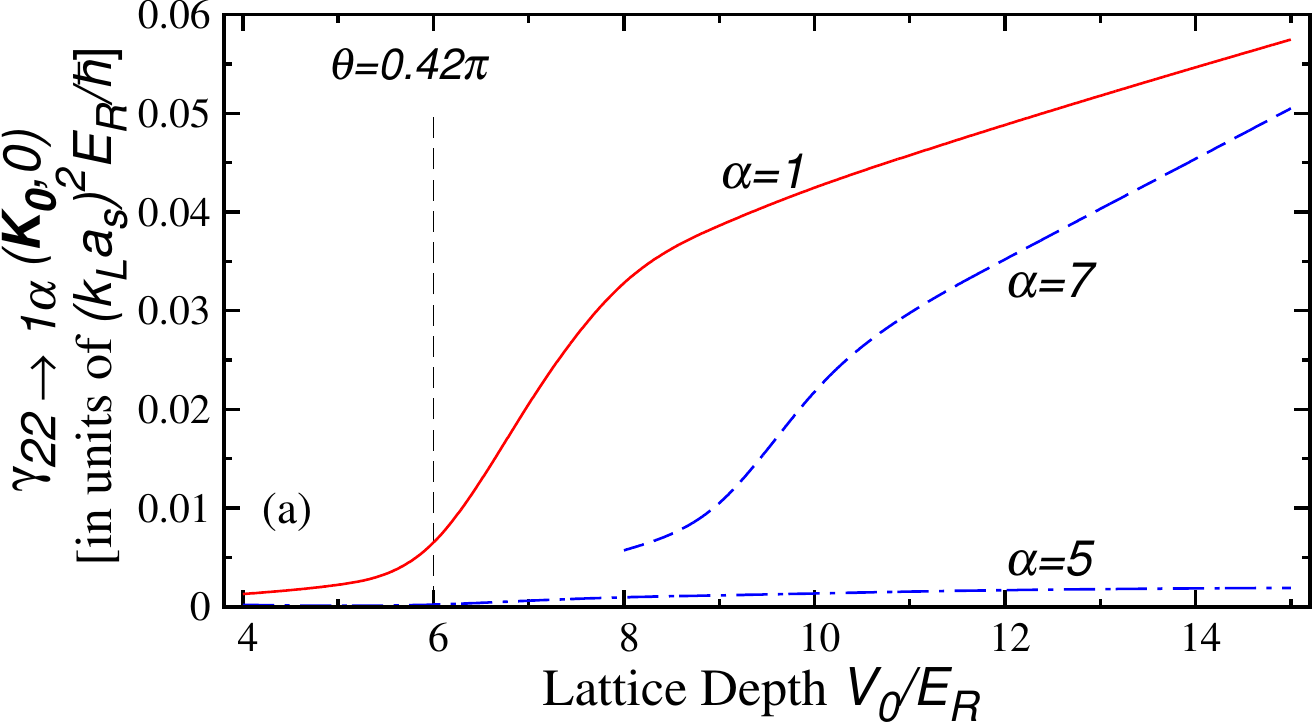} \label{decay-rate-vs-v}}\\
\subfloat{\includegraphics[width=3.3in]{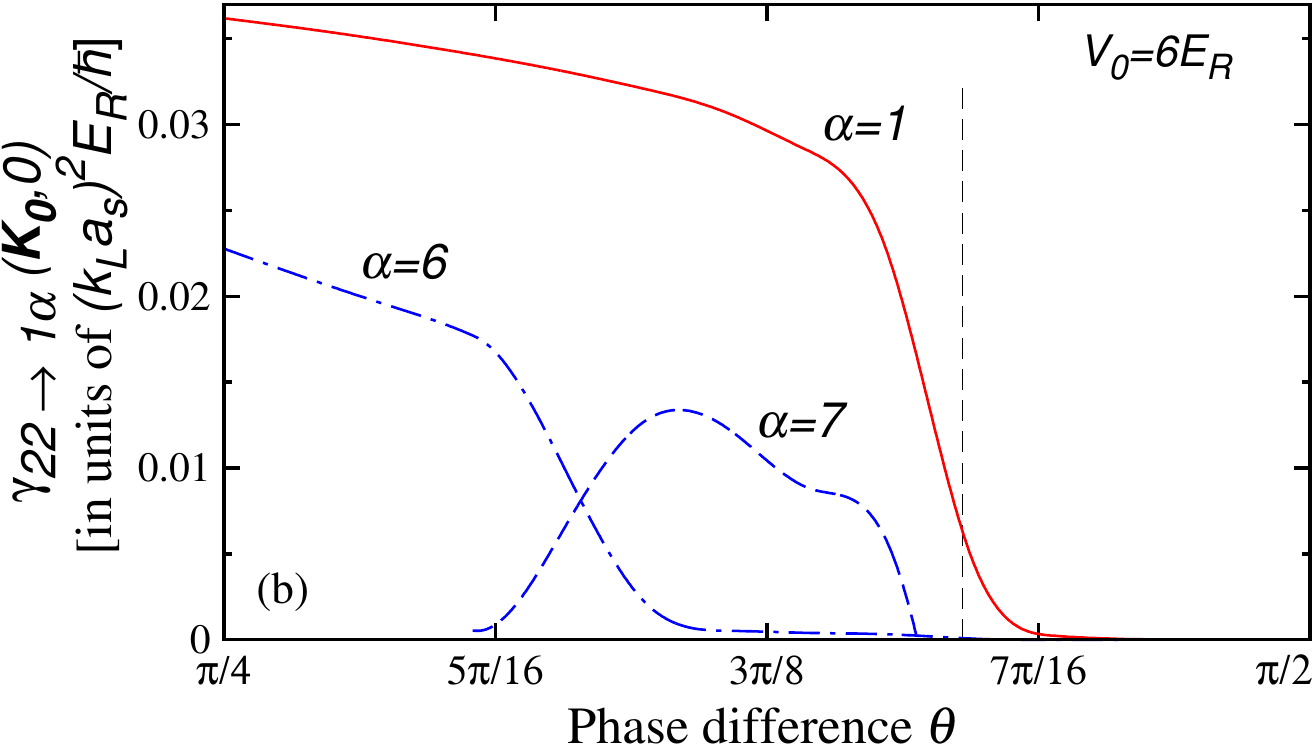} \label{decay-rate-vs-theta}}
\caption{(Color online) Decay rates $\gamma_{22 \to 1\alpha}(\K,0)$ for various bands $\alpha$, where two atoms are lost from the condensate, one going to the ground band and another to band $\beta$. 
Panel (a) shows the decay rates as a function of lattice depth $V_0/E_R$ at $\theta=0.42\pi$. The decay rate to band $\alpha=6$ is negligible and not shown in the figure. Decay processes involving still higher bands are energetically disallowed. The vertical dashed line indicates the lattice depth where the adjacent $s$ and $p$ orbitals are degenerate. Panel (b) shows the decay rates as a function of $\theta$ at lattice depth $V_0=6E_R$. Depending on the value of $\theta$, decay to higher bands $\alpha=6$ and $7$ becomes important. The vertical dashed line indicates the angle $\theta$ where the adjacent $s$ and $p$ orbitals are degenerate.} 
\label{decay-rates} 
\end{figure}  

After having evaluated the matrix elements, we are now ready to estimate the collisional loss rates for a condensed gas of atoms at temperature $T$ and total atom number $N$. This loss is dominated by pairs of atoms in the second band decaying to bands $\alpha=1$ and $\beta>4$. Following the discussion in Complement $\rm{C_I}$ in Ref.~\cite{Cohen_Tannoudji} and summing over all populated discrete states $\vert\Phi\rangle$, these rates per unit cell are given by
\begin{align}\label{gamma1}
  \Gamma_{22 \to 1\alpha}&=\frac{N_0(N_0-1)}{2M^2}\gamma_{22\to 1\alpha}(\K,0)\nonumber\\
           &\quad+\sum_{\mathbf{K},m'}\frac{N_0n(\mathbf K,m')}{M^2}\gamma_{22\to 1\alpha}(\mathbf K,m'),
\end{align}
where
\begin{align}\label{gamma2}
  \gamma_{22\to 1\alpha}(\mathbf K,m')=&\frac{E_R}{\hbar}(k_La_s)^2\frac{\pi\Gamma(m'+1/2)}{\Gamma(m'+1)}\nonumber\\
  &\times M\sideset{}{'}\sum_{\mathbf K_1,\mathbf K_2} (P_{22}^{1\alpha}(\mathbf K_1,\mathbf K_2,\mathbf K,\mathbf K_0)/k_L^2)^2\nonumber\\
  &\times\frac{\Gamma(-\delta/(2\hbar\omega_z)+1/2)}{\Gamma(-\delta/(2\hbar\omega_z)+1)},
\end{align}
$\delta=\epsilon_{1}(\mathbf K_1)+\epsilon_{\beta}(\mathbf K_2)-\epsilon_{2}(\mathbf K_0)-\epsilon_{2}(\mathbf K)$, and the prime in the momentum sum indicates that it has to conserve quasi-momentum. The rates $\gamma_{22\to 1\alpha}(\mathbf{K},m')$ are independent of $M$ and sums over harmonic oscillator indices have been performed analytically using the approximate form of $O_{m'0}^{mn}$.

The first term in Eq.~\eqref{gamma1} describes loss processes where two condensed atoms at quasi momentum $\K$ and harmonic oscillator level $m'=0$ are lost. The second term describes sum of loss processes where one condensed atom and another non-condensed atom at quasi momentum ${\mathbf K}$ and harmonic oscillator level $m'$ from the second band is lost. 

 For the decay process where both atoms are removed from the condensate and decay to the ground band $\alpha,\beta=1$, we can write down an analytical expression using the tight binding model.  For $\Delta>0$ it is
\begin{align}
  \gamma_{22\to 11}(\K,0)=\frac{V_0}{\hbar}(k_La_s)^2(1+\cos\theta)\frac{\sin^4\theta_{\K}}{16\sqrt{\pi}}\sqrt{\frac{\omega_z}{\omega_d}},\nonumber
\end{align}
where we have used $\delta\approx 2\hbar\omega_d$ and $\omega_z\ll\omega_d$. For $\Delta<0$, we have to replace $\sin\theta_{\K}$ by $\cos\theta_{\K}$.

\begin{figure}
\centering
\includegraphics[width=3.3in]{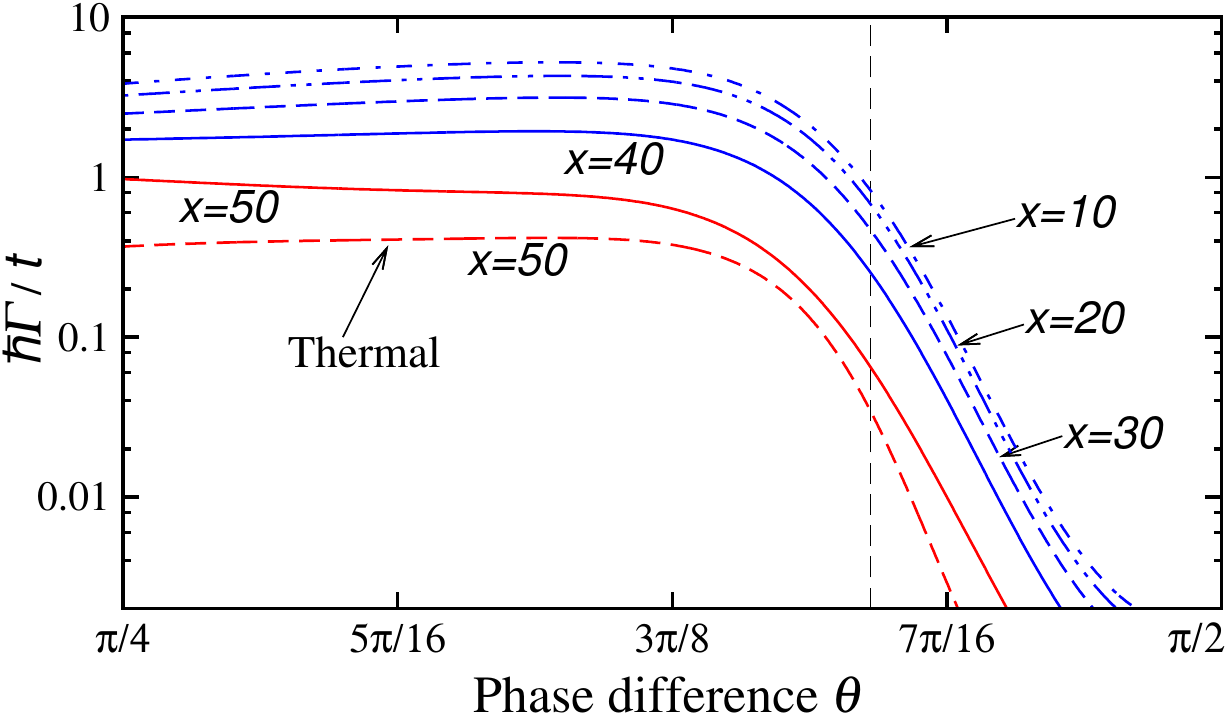}
\caption{(Color online) Dimensionless decay rate $\hbar\Gamma/t$ as a function of  phase difference $\theta$ for ${}^{87}{\rm Rb}$ atoms. We use lattice depth $V_0=6E_R$ and  atom number per lattice site of $\nu=200$. The different curves are at various values of dimensionless temperature $x=k_BT/\hbar\omega_z$. The vertical black dashed line represent $\theta=\theta_{res}$, the point at which the $s$ and $p$ orbitals of adjacent wells are degenerate. The dashed red curve shows the thermal contribution to the decay rate at $x=50$.}
\label{total-decay-vs-theta}
\end{figure}

Figure~\ref{decay-rate-vs-v} shows $\gamma_{22\to 1\alpha}(\K,0)$ based on Eq.~\eqref{gamma2} as a function of lattice depth $V_0/E_R$ at $\theta=0.42\pi$. The decay to the ground band is largest, followed by decay to the $7^{\it th}$ band, which is energetically allowed at lattice depths $V_0>8E_R$. For $V_0>8E_R$ within the tight binding model, the local orbitals $d_{x^2},d_{y^2}$ and $d_{xy}$ in the deep well form the basis for the $5^{\it th},6^{\it th}$ and $7^{\it th}$ bands. Moreover, only the $7^{\it th}$ band has significant $d_{xy}$ character at $\theta=0.42\pi$. We can then show that decay contributions to these three bands are significant only when the final state involves $d_{xy}$  orbitals and, hence, only the $7^{\it th}$ band has significant losses. Also, with increasing lattice depth, the harmonic-oscillator length in the deep well increases, leading to a much tighter confinement of the atoms. Thus, the decay rate increases with increasing lattice depths.

Figure~\ref{decay-rate-vs-theta} shows $\gamma_{22\to 1\alpha}(\K,0)$ as a function of the phase difference $\theta$ at lattice depth $V_0=6E_R$. Decay to the ground band is again the dominant term, followed by contribution to the $6^{\it th}$ and $7^{\it th}$ bands, depending on the phase difference $\theta$.  The steep decrease in the $\alpha=1$ decay rate  near $\theta=\theta_{res}$ with increase of $\theta$ is due to the change in character of the condensate wave function from predominantly $p$ state in the deep well to $s$ state in the shallow well. (The sharp increase of the $\alpha=1$ rate in Fig.~\ref{decay-rate-vs-v} has a similar origin.) The behavior of the loss rates of the $6^{\it th}$ and $7^{\it th}$ bands can be understood from studying Figs.\ref{onsite_energy} and \ref{bandgap-v6}. Between $\theta=\pi/4$ and $3\pi/8$ the three $d$ orbitals in the deep wells and the $s$ orbital in the shallow well are in resonance near $\theta=0.3\pi$ with tunneling energies that are of order of the  splittings. For $\theta<0.3\pi$, the $6^{\it th}$ band has predominant $d_{xy}$ character and is the only band that has a significant loss rate. For $\theta>0.3\pi$, it is the $7^{\it th}$ band that has significant $d_{xy}$ character. For $\theta>0.4\pi$, loss to $\alpha=7$ is not energetically allowed. Also with decreasing $\theta$, the harmonic-oscillator length in the deep well increases, and thereby the decay rate increases with decreasing $\theta$.

The total decay rate $\Gamma$ is obtained by summing over contributions from all bands $\alpha$. To a good approximation, this is given by
\begin{align}\label{decay-total}     \Gamma&\approx\sum_{\alpha}\left(\frac{\nu^2}{2}+\nu\frac{k_BT}{\hbar\omega_z}\sqrt{\frac{\pi\hbar\omega_z}{t/\F}}\right)\gamma_{22\to 1\alpha}(\K,0),
\end{align}
where $\nu=N/M$, the total number of atoms per unit cell. The above result is reached by first performing the sums over $\mathbf{K}_1$ and $\mathbf{K}_2$ in Eq.~\ref{gamma2} neglecting the quasi momentum dependence of the first and second band in $\delta$, since their band widths are negligible compared to the band gaps.  We then find
\begin{eqnarray}
\gamma_{22\to 1\alpha}(\mathbf{K},m')&=&\gamma_{22\to 1\alpha}(\K,0)\frac{\Gamma(m'+1/2)}{\sqrt{\pi}\Gamma(m'+1)}\\
   && \times \frac{\sin^2\theta_{\mathbf{K}}(\sin(K_xa)+\sin(K_ya))^2}{\sin^2\theta_{\K}(\sin(K_{\mathbf{0}x}a)+\sin(K_{\mathbf{0}y}a))^2},\nonumber
\end{eqnarray}
which separates its $\mathbf{K}$ and $m'$ dependence. Noting that $n(\mathbf{K},m')=\sum_{k=1}^{\infty}z^k\exp[-k\beta(\epsilon_2(\mathbf{K})+m'\hbar\omega_z)]$, we can perform the sum over $m'$ analytically. The sum over $\mathbf{K}$ can be performed in a manner similar to that used in determining the total atom number in section~\ref{Thermodynamics} and noting that $\epsilon_2(\mathbf{K})$ is an even function of both $K_x$ and $K_y$ so that only the even part of $\gamma_{22\to 1\alpha}(\mathbf{K},m')$ contributes to the sums. The energy $\epsilon_2(\mathbf{K})$ is Taylor expanded as before around $\mathbf{K}=\K$ and the exponential $\exp[-k\beta(\epsilon_2(\mathbf{K})]$ separated into a momentum dependent and  independent part leading to an integral that is solved to  $O([\beta t/\F]^2)$. Finally, we find the result of Eq.~\ref{decay-total}. 

Figure~\ref{total-decay-vs-theta} shows the total decay rate in units of the tunneling energy defined in section~\ref{TB-model} as a function of the phase difference $\theta$ for various temperatures below the critical temperature $T_c$. The red curve is at $k_BT=50\hbar\omega_z$, which is closest to the critical temperature. For $\theta<\theta_{res}$, the loss rates are large but weakly dependent on $\theta$ while for large $\theta$ the loss rates rapidly decrease. Qualitatively the curves follow the behavior of $\alpha=1$ loss rates shown in Fig.~\ref{decay-rate-vs-theta}. The slight deviations between the curves at different temperatures are due to a non-trivial redistribution of atoms between the condensate and the thermal component. Figure~\ref{total-decay-vs-theta} also shows the loss rate from the thermal atoms for $k_BT/\hbar\omega_z=50$. Near $\theta=\theta_{res}$, it corresponds to about $50\%$ of the total loss. For smaller temperatures, the contribution from the thermal atoms becomes smaller.

We want to make a final observation about the use of the tight binding model. The decay rates involving bands $\alpha=3$ and $4$ are zero both in the tight binding approximation and for the exact evaluation of their wavefunctions. Hence, the fact that the tight binding model is insufficient for bands $\alpha=3$ and $4$ does not affect the decay rates. For bands $\alpha\ge 5$, we have used the numerical wavefunctions.  
\end{section}

\begin{section}{Conclusion}\label{Conclusion}
  We have studied two aspects of Bose-Einstein condensates formed in the second band of a 2 dimensional optical lattice with weak harmonic confinement in the third direction. The first aspect relates to the lattice transformation process that leads to atom population in excited bands, and the subsequent thermalization process in the excited bands. A non-interacting Bose gas picture provides analytic expressions for the condensed fraction of atoms and the corresponding critical temperature $T_c$ as a function of lattice parameters reached after the lattice transformation. These crucially depend on $\Delta$, the detuning between the adjacent $s$ and $p$ orbitals. The analysis shows that $k_BT_c>t\gg\hbar\omega_z$ and the condensed fraction is minimal at $\Delta=0$. We also show that the lattice transformation process, in general, leads to a heating of the atom cloud. For large positive detuning $\Delta$, this heating is significant, and the final temperature $T_f\gg T_c$. At $\Delta=0$, which is the case of interest since the adjacent $s$ and $p$ orbitals are degenerate at this detuning, even for $T_i\approx 0$, $T_f$ is large but still just below the corresponding $T_c$. Thus, after thermalization, the condensed fraction in the excited band is small.

The second aspect deals with the lifetime of the condensate formed in the second band of the optical lattice, which is  determined by atom-atom elastic collisions. All such decay processes only involve the deep wells, since one or both of the colliding atoms decay to quasi momentum states in the ground band, which in the tight binding picture are predominantly determined by the localized ground states in the deep well. The available oscillator states along the $z$ axis ensure that for all lattice parameters, there exist a  dominant decay process in which two atoms from the second band, undergoing elastic collisions, both decay to the ground band. The excess energy is released into excitations in the $z$ oscillator levels. At detuning $\Delta\le 0$, the above process is then the only energetically-allowed decay process. We provide analytical results for this process as function of lattice parameters, atom number and temperature using the tight binding model. We show that the contribution to this decay from thermal atoms in the second band is significant. For $\Delta>0$ and depending on lattice depth, other channels involving the $5^{\it th}$, $6^{\it th}$ or $7^{\it th}$ band contribute to the decay process. These decay rates are determined using tight binding results for bands $\alpha\le 4$ and numerical results for bands $\alpha>4$. We show that at temperatures close to $T_c$, the total decay rate $\Gamma\ll t/\hbar$, the tunneling rate between adjacent $s$ and $p$ orbitals. However, the total decay rate becomes comparable to or larger than the tunneling rate  for larger detuning $\Delta$ and $T\ll T_c$.

In the future, we would like to address two other aspects of dynamics of atoms in  this lattice geometry. 
The first aspect is the thermalization time scale for atoms in the excited bands after the lattice transformation. This should prove to be an interesting study, particularly comparing this to other time scales in the problem, and studying this for different lattice transformations. The other aspect of the lattice is that the first three excited bands intersect at the center of the first Brillouin zone, with the second and fourth bands forming a Dirac cone. This point of intersection has topological significance. Extending the present study to atoms excited to still higher bands, particularly atoms excited to the $4^{\it th}$ band where they condense at the center of the first Brillouin zone will be quite interesting.
\end{section}

\bibliography{references-paper}
\end{document}